\documentclass[twocolumn]{aastex63}

\pdfoutput=1

\usepackage{natbib}
\usepackage{color}
\usepackage{bm}
\usepackage{hyperref}
\usepackage{float}
\usepackage{latexsym}
\usepackage{amsmath}
\usepackage{amssymb}
\usepackage{multirow}
\usepackage{rotate}
\usepackage{graphicx}
\usepackage{url}
\usepackage{color}
\usepackage{afterpage}

\submitjournal{ApJ}
\accepted{\today}

\begin{document}

	\title{The Mass and Age Distribution of Halo White Dwarfs in the Canada-France Imaging Survey}

	\shorttitle{The Mass and Age Distribution of High Velocity White Dwarfs}
	\shortauthors{Fantin et al.}
	
	\smallskip

	\author[0000-0003-3816-3254]{Nicholas J. Fantin}
	\correspondingauthor{Nicholas J. Fantin}
	\email{nfantin@uvic.ca}
	\affil{Department of Physics and Astronomy, University of Victoria, Victoria, BC, V8P 1A1, Canada}
	\affil{National Research Council of Canada, Herzberg Astronomy \& Astrophysics Research Centre, 5071 W. Saanich Rd, Victoria, BC, V9E 2E7, Canada}
	
	\author{Patrick C{\^o}t{\'e}}
	\affil{National Research Council of Canada, Herzberg Astronomy \& Astrophysics Research Centre, 5071 W. Saanich Rd, Victoria, BC, V9E 2E7, Canada}
	
	\author{Alan W. McConnachie}
	\affil{National Research Council of Canada, Herzberg Astronomy \& Astrophysics Research Centre, 5071 W. Saanich Rd, Victoria, BC, V9E 2E7, Canada}

	\author{Pierre Bergeron} 
	\affil{D{\'e}partement de Physique, Universit{\'e} de Montr{\'e}al, C.P. 6128, Succ. Centre-Ville, Montr{\'e}al, QC H3C 3J7, Canada}
	
	\author{Jean-Charles Cuillandre}
	\affil{AIM, CEA, CNRS, Universit{\'e} Paris-Saclay, Universit{\'e} Paris Diderot, Sorbonne Paris Cit{\'e}, Observatoire de Paris, PSL University, F-91191 Gif-sur-Yvette Cedex, France}
	
	\author{Patrick Dufour} 
	\affil{D{\'e}partement de Physique, Universit{\'e} de Montr{\'e}al, C.P. 6128, Succ. Centre-Ville, Montr{\'e}al, QC H3C 3J7, Canada}
	
	\author{Stephen D. J. Gwyn}
	\affil{National Research Council of Canada, Herzberg Astronomy \& Astrophysics Research Centre, 5071 W. Saanich Rd, Victoria, BC, V9E 2E7, Canada}
	
	\author{Rodrigo A. Ibata}
	\affil{1Universit{\'e} de Strasbourg, CNRS, Observatoire astronomique de Strasbourg, UMR 7550, F-67000 Strasbourg, France}
	
	\author{Guillaume F. Thomas \bigskip}
	\affil{Instituto de Astrof{\'i}sica de Canarias, E-38205 La Laguna, Tenerife, Spain}
	\affil{Universidad de La Laguna, Dpto. Astrof{\'i}sica, E-38206 La Laguna, Tenerife, Spain}

	\begin{abstract}
		We present optical spectroscopy for 18 halo white dwarfs identified using photometry from the Canada-France Imaging Survey and Pan-STARRS1 DR1 3$\pi$ survey combined with astrometry from \textit{Gaia} DR2. The sample contains 13 DA, 1 DZ, 2 DC, and two potentially exotic types of white dwarf. We fit both the spectrum and the spectral energy distribution in order to obtain the temperature and surface gravity, which we then convert into a mass, and then an age, using stellar isochrones and the initial-to-final mass relation. We find a large spread in ages that is not consistent with expected formation scenarios for the Galactic halo. We find a mean age of 9.03$^{+2.13}_{-2.03}$\,Gyr and a dispersion of 4.21$^{+2.33}_{-1.58}$\,Gyr for the inner halo using a maximum likelihood method. This result suggests an extended star formation history within the local halo population.
		\bigskip
		\bigskip\bigskip
		
	\end{abstract}
	
	\section{Introduction}
	\label{sec:intro}
	
	\begin{figure*}[!t]
		
		\includegraphics[angle=0,width=0.34\textwidth]{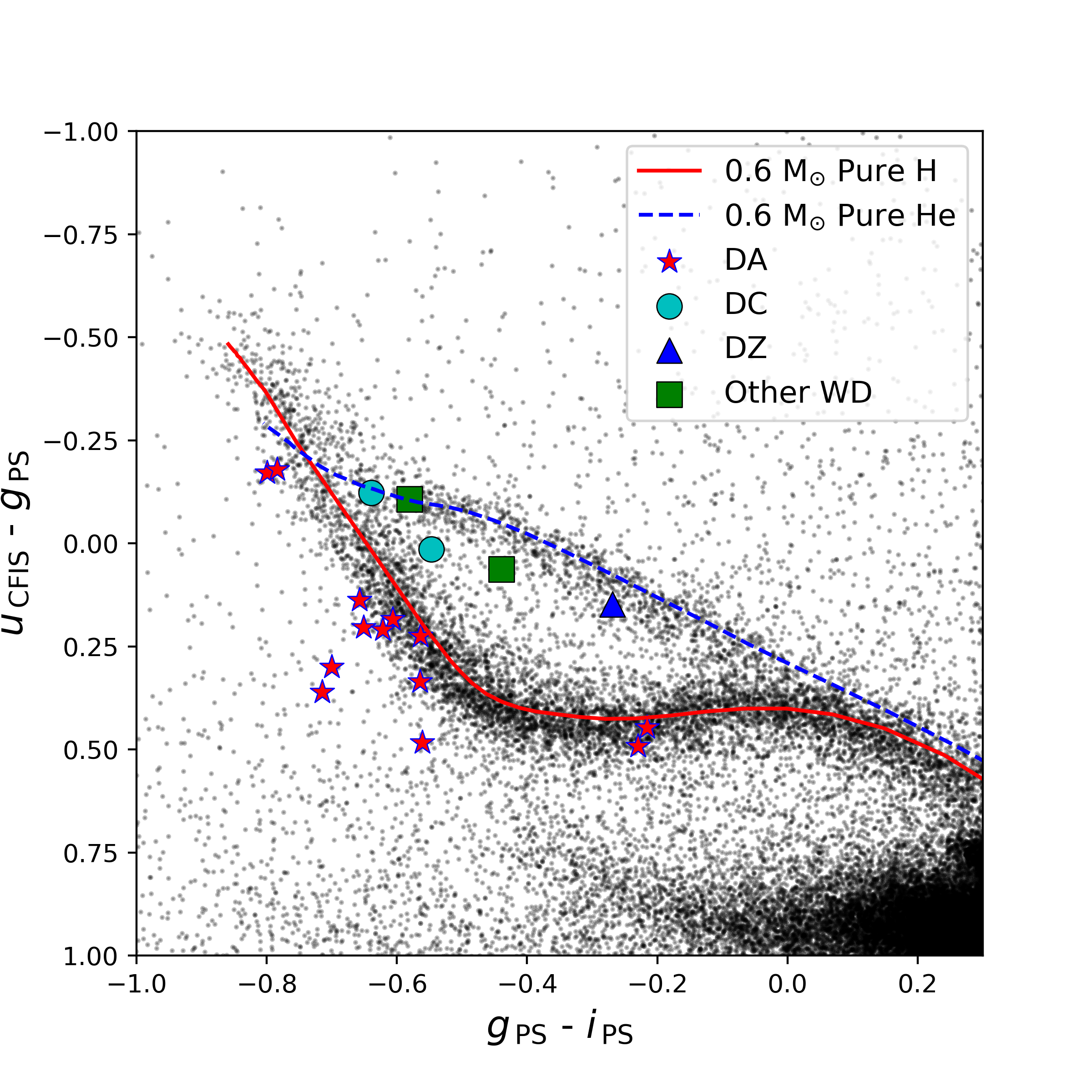}
		\includegraphics[angle=0,width=0.34\textwidth]{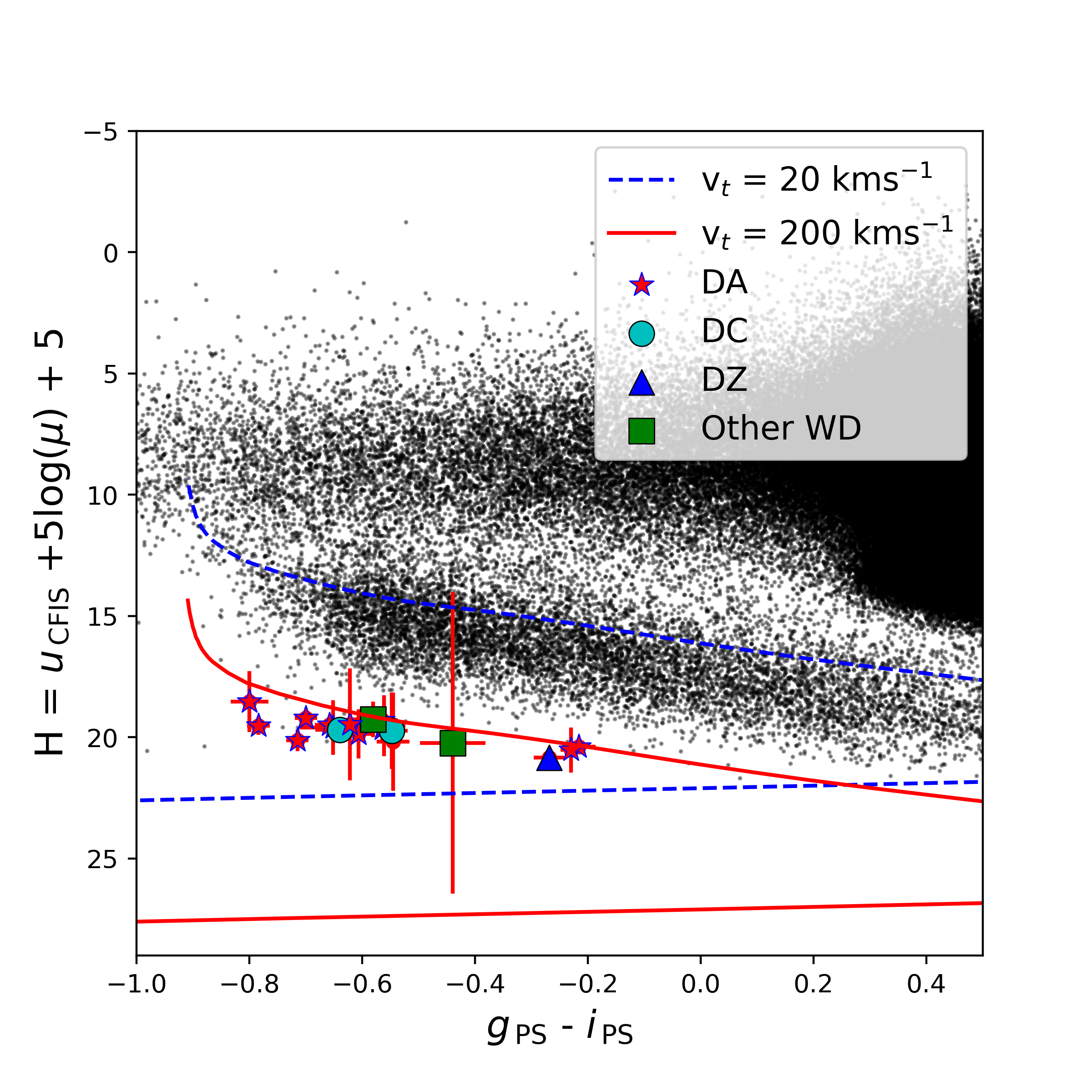}
		\includegraphics[angle=0,width=0.34\textwidth]{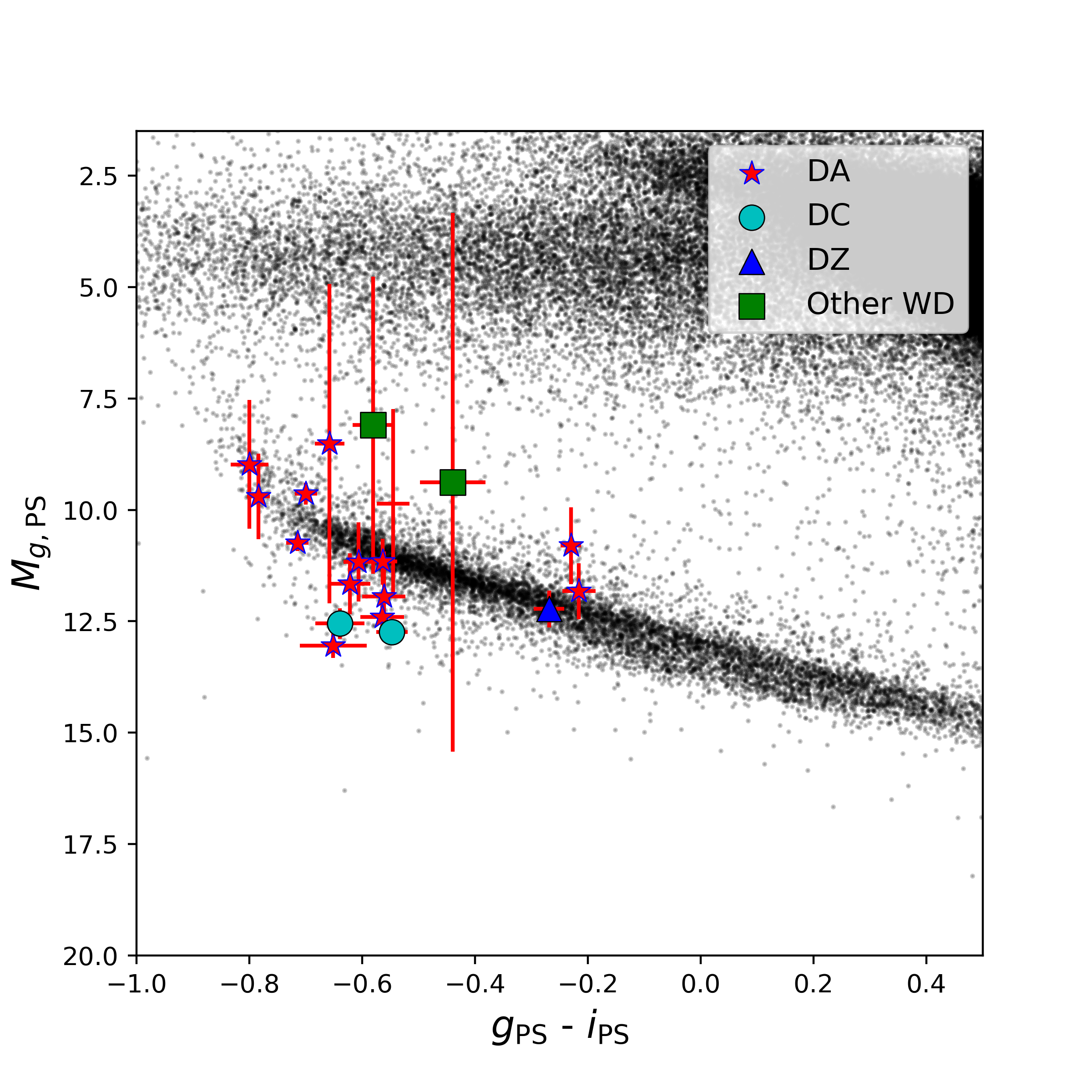}
		\caption{\textit{Left:} Colour-colour plot of all CFIS sources ($u <$ 21) with our 2019A halo sample highlighted based on their spectral type as determined in this work}. Red and blue lines represent model cooling tracks for 0.6\,M$_{\odot}$ white dwarfs with hydrogen and helium atmospheres respectively. \textit{Middle:} Reduced Proper Motion Diagram for our CFIS-PS1-\textit{Gaia} objects ($u <$ 21). Also included are model tracks for the thin disk ($v_{\textrm{t}}$ = 20\,kms$^{-1}$), and stellar halo populations ($v_{\textrm{t}}$ = 200\,kms$^{-1}$). Our halo white dwarf candidates were selected if their reduced proper motion was greater then the 200\,kms$^{-1}$ curve. \textit{Right:} Color-Magnitude diagram for the observed white dwarfs with symbols denoting their spectral type. Also plotted are objects from \textit{Gaia} DR2 to highlight the separation between the white dwarf cooling sequence and other point-sources. Distances are calculated using the \textit{Gaia} DR2 parallaxes.
		\bigskip 
		\label{fig:selection}
	\end{figure*}
	
	The Galactic halo contains the most ancient stars in the Milky Way, and as such has provided a means to study the early formation and evolution of our Galaxy. Many of these studies benefit from an accurate measurement of the age of a star, however, an age cannot be measured directly and instead relies on theoretical or semi-empirical models \cite{Soderblom2010}. One such star which has been used as a cosmochronometer is the white dwarf as its evolution is defined by a robust cooling curve. Their evolution is largely determined by their mass and atmospheric composition which can easily be measured either spectroscopically or photometrically, provided an accurate parallax is available \citep{Bergeron2005, Kalirai2012, Kilic2019}.

	Various methods using white dwarfs have been used to determine the ages of a variety of stellar populations, from the oldest globular clusters in the halo to younger open clusters in the disk. The age of the Milky Way has also been measured using field white dwarfs \citep[see, e.g,][]{Winget1987, Hansen2007, Bedin2009, Tremblay2014,Kilic2019}. Typically these studies rely on the white dwarf luminosity function as white dwarfs will bunch up at cool temperatures given the finite age of the population \cite{Harris2006, RowellHambly2011, Kilic2017}. These methods rely on the observations of the oldest white dwarfs, which are cooler and fainter than their younger counterparts, making them difficult to detect in optical surveys. 
	
	The Galactic halo, however, is also continuously producing young, hot, white dwarfs. These objects have lower masses than their disk counterparts as they evolve from older, lower mass, progenitors \citep{Kalirai2012, Cummings2018}. Measuring the age of these white dwarf has presented some unique challenges. Since the total age of the star is the sum of the white dwarf cooling age and the pre-white dwarf lifetime of its progenitor, both must be determined. The white dwarf cooling age can be determined from the mass, temperature, and atmospheric composition. The difficulty lies in measuring the progenitor lifetime, which dominates the total age of these young white dwarfs, as it relies on the ability to relate the observed white dwarf mass to the progenitor mass via the initial-to-final mass relation (IFMR).

	There exist discrepancies in the various IFMRs, however, much of the disagreement occurs at either high or low initial mass --- the most poorly sampled regions of parameter space. At low mass (important for the young Galactic halo white dwarfs) there exist very few nearby globular clusters for which spectroscopic white dwarf masses can currently be obtained. Currently, the low-mass IFMR is anchored by spectroscopic measurements of just a single globular cluster, Messier 4 (M4), performed by \cite{Kalirai2008}, which revealed white dwarf masses between 0.50 and 0.58 M$_{\odot}$. \cite{Kalirai2012} used the measured masses of four halo white dwarfs, combined with this IFMR to infer an inner halo age of (11.7\,$\pm$\,0.7)\,Gyr.

	Given that the age of the local halo remains uncertain given the small sample size acquired by \cite{Kalirai2012}, it is not surprising that the star formation history (SFH) of the halo also remains uncertain. In \cite{Fantin2019}, we concluded that the photometric sample of halo objects in our dataset was insufficient, particularly relative to the disk, to measure an accurate halo SFH. One solution is to acquire spectroscopic temperatures and masses in order to directly measure the age distribution of a sample of halo white dwarfs.  
	
	In this paper, we build on the sample of \cite{Kalirai2012} by acquiring Gemini-GMOS spectroscopic observations of 18 halo white dwarfs. In Section \ref{sec:data} we detail our selection method and all photometric and spectroscopic observations. We present the resulting masses and ages in Section \ref{sec:Results},  discuss their implications in Section \ref{sec:halo_discussion}, before summarizing and looking towards the future in Section \ref{sec:conclusions}.

	\section{Data}
	\label{sec:data}
	
	\subsection{Photometric Data}
	
	The data used as part of this study were acquired as part of the Canada France Imaging Survey \citep[CFIS;][]{CFIS1}, Pan-STARRS1 DR1 3$\pi$ \citep[PS1;][]{PS1}, and \textit{Gaia} DR2 \citep{Gaia}. CFIS is a photometric survey comprised of $\sim$4,900\,deg$^2$, out of an eventual 10,000\,deg$^2$, of \textit{u}-band data 3 magnitudes fainter than the Sloan Digital Sky Survey (SDSS) at equivalent uncertainties. CFIS was then combined with PS1 \textit{grizy} photometry and \textit{Gaia} DR2 astrometry in the same manner as \cite{Thomas2018}. A sample color-color diagram using the CFIS \textit{u}-band combined with the PS1 \textit{g}- and \textit{i}-bands can be seen in the left-hand panel of Figure \ref{fig:selection}, showing the clean sequence of white dwarfs, QSOs, and other stellar objects --- a result of the exquisite photometry from both surveys. 
	
	\subsection{Sample Selection}

	In this paper, we target the white dwarfs with the most extreme tangential velocities --- those most likely associated with the halo. Our targets were selected based on their location in a reduced proper motion diagram (RPMD; see the middle panel of Figure \ref{fig:selection}). The reduced proper motion, $H$, can be related to the absolute magnitude, $M$, and tangential velocity, $v_{\textrm{t}}$, as follows:
	
	\begin{equation}
		\label{equation:RPM}
		\begin{array}{lcl}
			H & = & m + 5\log\mu + 5 \\
			& = & M + 5\log v_{\textrm{t}} - 3.379. \\
		\end{array}
	\end{equation}
	
	\begin{figure*}[!t]
		
		\includegraphics[angle=0,width=\textwidth]{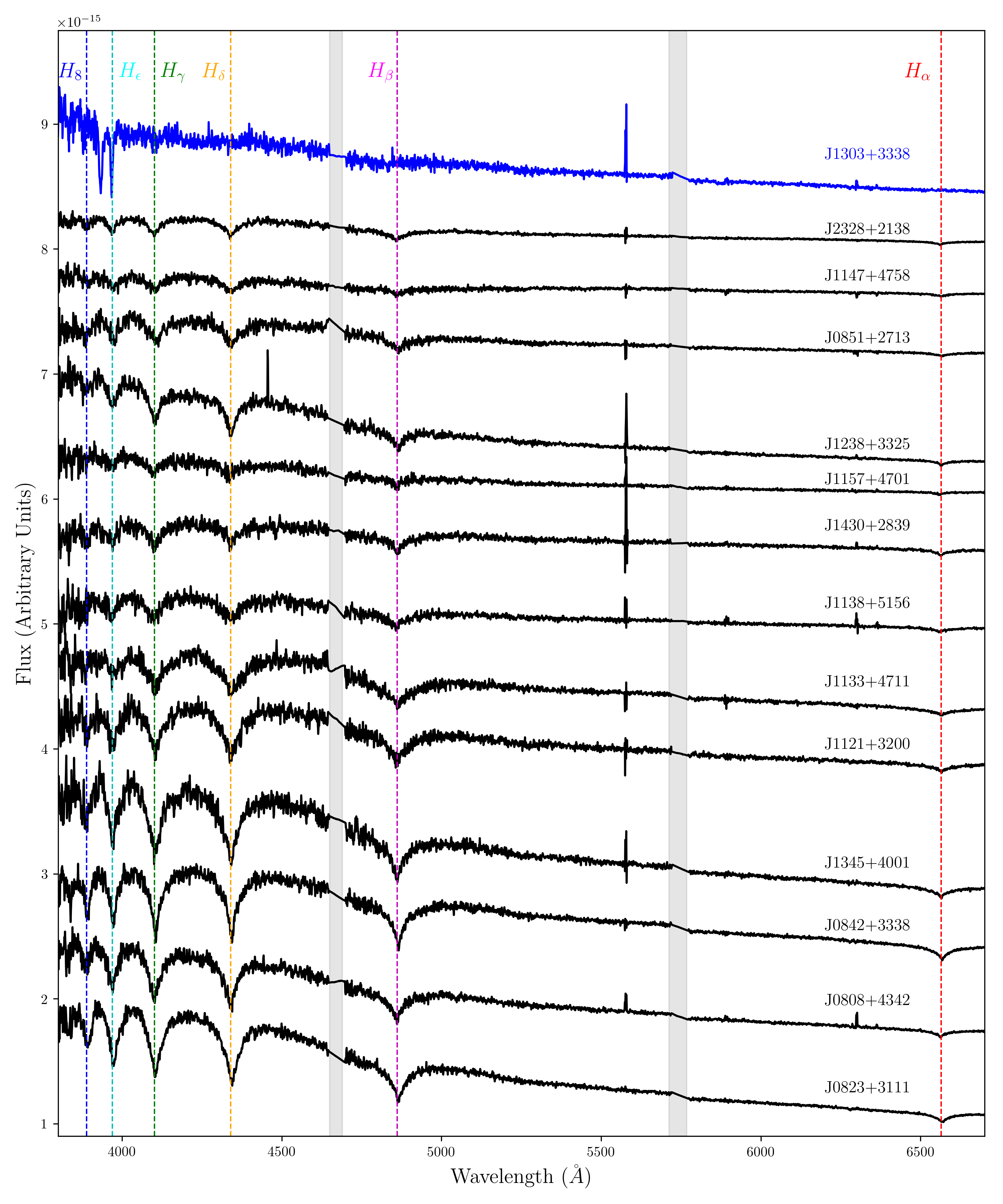}
		
		\caption{Spectra of 14 high-velocity white dwarfs obtained using GMOS on Gemini-North for which a spectroscopic mass can be determined. The top spectrum, J1303-3338, is a DZ with prominent Ca H \&K absorption, while the remaining are classified as DA. The Balmer series is marked for clarity by the dashed lines. The chip gaps on the GMOS detector are located within the shaded grey regions. 
			\bigskip }
		\label{fig:spectra}
	\end{figure*}
	
	\begin{figure*}[!t]
		
		\includegraphics[angle=0,width=\textwidth]{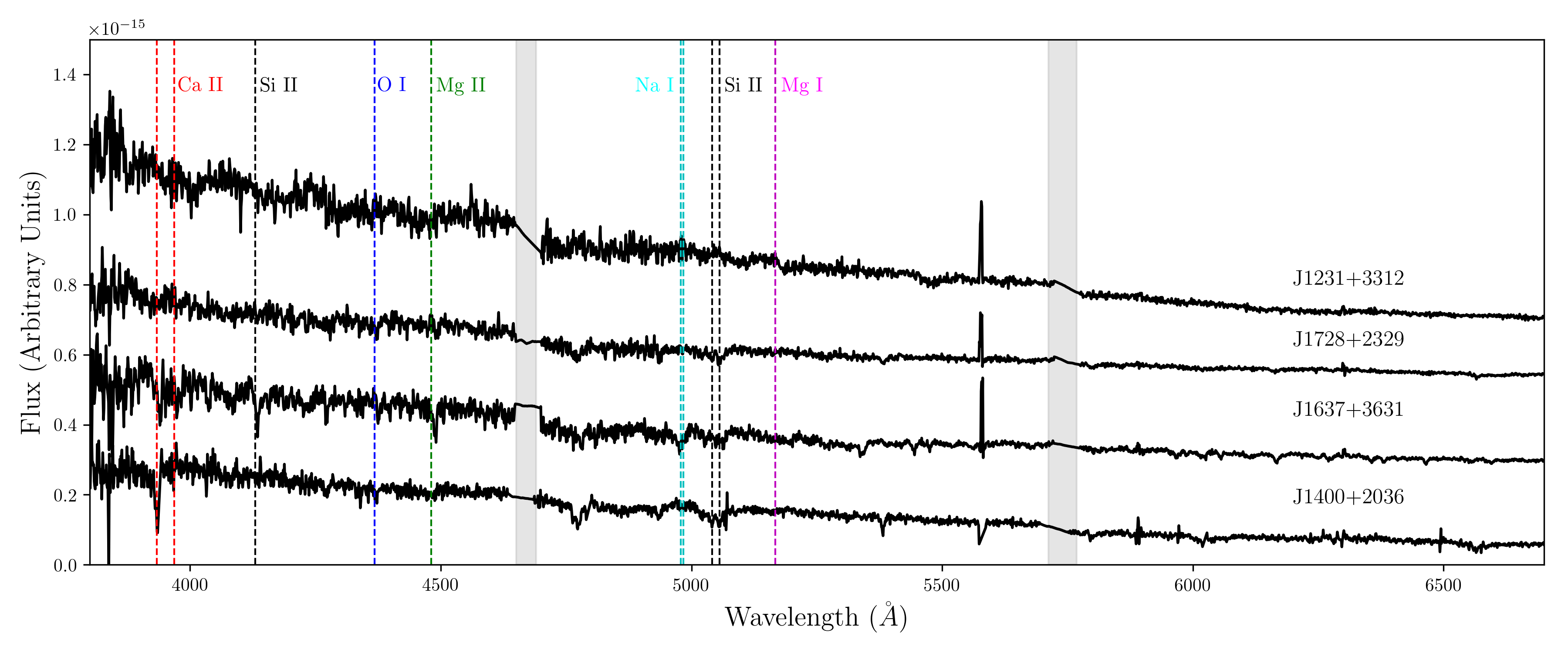}
		
		\caption{GMOS spectra of the four non-DA or DZ white dwarfs.  J1637+3631, a remnant of a peculiar thermonuclear reaction presented in detail by \cite{Raddi2019}, is shown as the second from the bottom. A number of common absorption features are highlighted to guide the eye. The chip gaps on the GMOS detectors are located within the shaded grey regions as in Figure \ref{fig:spectra}.
			\bigskip 
			\bigskip}
		\label{fig:contamination}
	\end{figure*}
	
	\noindent where $m$ is the apparent magnitude and $\mu$ is the proper motion in arcseconds per year. Given that white dwarfs are intrinsically fainter than their main-sequence counterparts at equal temperatures they will experience a larger proper motion and end up with larger reduced proper motion values. This method can cleanly separate white dwarfs from other point-sources and previous studies, for example, have found a contamination rate ranging from 1-5\,\% from halo sub-dwarfs \citep[see, e.g,][]{Kilic2006,Dame2016}.
	
	In the middle panel of Figure \ref{fig:selection} we have plotted model cooling tracks with tangential velocities of 20\,kms$^{-1}$ and 200\,kms$^{-1}$, which are typically associated with the mean tangential velocity of the thin disk and stellar halo respectively. The synthetic magnitudes have been calculated in the CFIS-\textit{u} and PS1 \textit{grizy} bands for H- and He-atmosphere models using the procedure described in \citet{HolbergBergeron2006}. The white dwarf cooling sequences are similar to those described in \citet{Fontaine2001} with (50/50) C/O-core compositions, $M_{\rm He}/M_{\star}=10^{-2}$, and $M_{\rm H}/M_{\star}=10^{-4}$ or $10^{-10}$ for H- and He-atmosphere white dwarfs, respectively \footnote{See \url{http://www.astro.umontreal.ca/~bergeron/CoolingModels}}.
	
	The CFIS white dwarf catalog contains 25,156 candidates below the 20\,kms$^{-1}$ cooling curve over $\sim$4,900\,deg$^2$ of the Northern sky \citep{Fantin2019}. We consider an object to be part of the Galactic halo if it lies below the model halo track (i.e $v_{\textrm{t}} >$ 200\,kms$^{-1}$). Applying this selection results in 90 halo white dwarf candidates, however, we target the 17 brightest objects that were observable during the 2019A semester from Gemini-North and the brightest object observable with Gemini-South. These objects all have $u$-band magnitudes brighter than 21 AB mag, and half of the objects are brighter than $u$ = 20 AB mag. Photometric data associated with the objects observed are presented in Table \ref{table:observations}.

	\begin{table*}[!t]
		\centering
		\caption{Halo white dwarfs observed with GMOS}
		\begin{tabular}{cccccccccccc}
			\hline
			RA      & Dec   & \textit{u}$_{\textrm{CFIS}}$  & $g_{\textrm{PS}}$       & $i_{\textrm{PS}}$    & $\mu_{\alpha}$          &$\mu_{\delta}$ & parallax \\

			HH:MM:SS& HH:MM:SS   & AB mag  & AB mag      & AB mag   &  $^{\prime\prime}$/yr       &$^{\prime\prime}$/yr & mas \\
			
			\hline
			08:08:48.96&43:42:14.40&20.04\,$\pm$\,0.01&20.04\,$\pm$\,0.02&20.30\,$\pm$\,0.02&$-$4.5\,$\pm$\,1.5&$-$77.1\,$\pm$\,0.8 & 0.49\,$\pm$\,0.82 \\
			
			08:23:59.15&31:11:52.10&18.31\,$\pm$\,0.00&18.11\,$\pm$\,0.00&18.81\,$\pm$\,0.02&$-$15.4\,$\pm$\,0.3&$-$165.5\,$\pm$\,0.2 & 2.02\,$\pm$\,0.23 \\
			
			08:42:49.37&33:38:43.50&20.22\,$\pm$\,0.01&19.81\,$\pm$\,0.01&20.04\,$\pm$\,0.01&$-$67.4\,$\pm$\,0.9&$-$121.3\,$\pm$\,0.5& 1.58\,$\pm$\,0.63\\
			
			16:37:12.21&36:31:55.90&19.99\,$\pm$\,0.01&20.14\,$\pm$\,0.02&20.71\,$\pm$\,0.02&$-$18.5\,$\pm$\,1.2&64.3\,$\pm$\,1.2 & 0.39\,$\pm$\,0.60  \\
			
			08:51:15.42&27:13:54.30&19.69\,$\pm$\,0.01&19.47\,$\pm$\,0.01&20.04\,$\pm$\,0.01&$-$21.1\,$\pm$\,0.8&$-$90.0\,$\pm$\,1.6 & 2.17\,$\pm$\,0.52\\
			
			11:21:31.34&32:00:12.30&20.27\,$\pm$\,0.01&19.86\,$\pm$\,0.01&20.42\,$\pm$\,0.04&36.6\,$\pm$\,1.4&$-$77.6\,$\pm$\,2.1& 2.61\,$\pm$\,1.21 \\

			11:33:27.50&47:11:36.50&19.87\,$\pm$\,0.01&19.73\,$\pm$\,0.02&20.38\,$\pm$\,0.05&$-$73.4\,$\pm$\,0.7&$-$59.6\,$\pm$\,0.9 & 4.62\,$\pm$\,0.57 \\
			
			11:38:59.14&51:56:47.70&20.34\,$\pm$\,0.01&20.04\,$\pm$\,0.02&20.60\,$\pm$\,0.03&27.4\,$\pm$\,0.8&$-$79.3\,$\pm$\,0.7& 2.98\,$\pm$\,0.62 \\
			
			11:47:55.58&47:58:13.80&20.13\,$\pm$\,0.01&20.02\,$\pm$\,0.01&20.62\,$\pm$\,0.02&$-$82.2\,$\pm$\,0.5&$-$43.8\,$\pm$\,0.8 & 1.70\,$\pm$\,0.70\\
			
			11:57:51.05&47:01:23.10&19.32\,$\pm$\,0.01&19.57\,$\pm$\,0.01&20.35\,$\pm$\,0.02&$-$32.5\,$\pm$\,0.5&$-$92.9\,$\pm$\,0.5& 1.06\,$\pm$\,0.47\\
			
			12:31:25.29&33:12:13.10&19.51\,$\pm$\,0.01&19.51\,$\pm$\,0.02&20.06\,$\pm$\,0.02&$-$105.4\,$\pm$\,0.8&33.3\,$\pm$\,0.5 & 4.41\,$\pm$\,0.46\\
			
			12:38:05.70&35:25:23.70&19.15\,$\pm$\,0.01&19.36\,$\pm$\,0.01&20.16\,$\pm$\,0.03&$-$57.8\,$\pm$\,0.7&$-$36.0\,$\pm$\,0.5& 0.84\,$\pm$\,0.56  \\
			
			13:03:34.31&33:38:43.30&20.09\,$\pm$\,0.01&20.04\,$\pm$\,0.03&20.30\,$\pm$\,0.02&$-$3.9\,$\pm$\,1.0&$-$145.2\,$\pm$\,0.9& 2.74\,$\pm$\,0.52 \\
			
			13:45:51.89&40:01:00.90&18.73\,$\pm$\.0.01&18.63\,$\pm$\,0.01&19.34\,$\pm$\,0.02&$-$192.4\,$\pm$\,0.2&$-$52.8\,$\pm$\,0.3& 2.65\,$\pm$\,0.22 \\
			
			17:28:02.89&23:29:12.20&19.90\,$\pm$\,0.01&20.02\,$\pm$\,0.03&20.66\,$\pm$\,0.03&6.8\,$\pm$\,0.6&$-$85.6\,$\pm$\,0.8 & 3.20\,$\pm$\,0.51 \\
			
			14:30:32.55&28:39:11.10&20.44\,$\pm$\,0.01&20.03\,$\pm$\,0.03&20.25\,$\pm$\,0.02&$-$10.6\,$\pm$\,1.3&$-$117.0\,$\pm$\,1.4& 2.28\,$\pm$\,0.66   \\
			
			23:28:55.81&21:38:26.4&20.05\,$\pm$\,0.01&19.92\,$\pm$\,0.01&20.54\,$\pm$\,0.03&76.2\,$\pm$\,1.1&$-$27.8\,$\pm$\,0.8& 2.22\,$\pm$\,0.69   \\
			
			\hline
			
			\tablenotemark{a}14:00:11.93&20:36:54.40&20.45\,$\pm$\,0.01&20.56\,$\pm$\,0.03&21.00\,$\pm$\,0.05&$-$85.8\,$\pm$\,2.9&1.3\,$\pm$\,3.9& 0.58\,$\pm$\,1.61  \\
			\hline
		\end{tabular}
		\label{table:observations}
		\tablenotetext{a}{J1400+2036 was observed using Gemini South}
		\bigskip
		\bigskip
	\end{table*}

	\subsection{Spectroscopic Follow-up}

	Long-slit spectroscopic follow-up observations were performed using the Gemini Multi-Object Spectrographs (GMOS) at the twin 8-m Gemini Observatories on Mauna Kea and Cerro Pach{\'o}n. To achieve the desired spectral range, the B600 grating with a central wavelength of 5200\,\AA was used. This allowed for the simultaneous acquisition of H$_{\alpha}$ through to H$_{\textrm{8}}$, which is ideal for mass and temperature determinations. A focal plane unit (slit width) of 0\farcs75 was selected in order to maximize the SNR while maintaining a suitable resolution. The data was binned 2x2, resulting in a resolution of R$\sim$1200.
	
	Using the recommendation from \cite{Kepler2006} we set our exposure times as to acquire spectra with a signal-to-noise ratio (SNR) of 15 per resolution element at the location of the higher order Balmer lines (H8) which are crucial for untangling the degeneracy between surface temperature and surface gravity near 12,000\,K, a region where some of our targets were expected to lie. The resulting integration times were split into sub-exposures in order to minimize the impact of cosmic rays. Given that we were awarded Band 2 time, however, the variable observing conditions resulted in a variety of observed SNR values (see Table \ref{table:fits_gemini}).

	\begin{table*}[!t]
		\centering
		\caption{Properties from Spectral and Photometric Fitting}
		\begin{tabular}{l|cc|ccc|ccc}
			\hline
			&&&&Spectroscopy&&&Photometry\tablenotemark{b}&\\
			\hline
			ID & SNR\tablenotemark{a} &Type & T$_{\textrm{eff}}$(K)   & Mass  (M$_{\odot}$)& t$_{cool}$ (Myr)& T$_{\textrm{eff}}$(K)   & Mass  (M$_{\odot}$) & t$_{cool}$ (Myr)  \\
			\hline
			
			J0808+4342 & 15.6 &DA& 22206\,$\pm$\,543&0.531\,$\pm$\,0.036& 28 & 21209\,$\pm$\,1122 & 0.194$_{-0.283}^{+0.41}$ & 74\\
			
			J0823+3111 &20.2 &DA& 19763\,$\pm$\,366& 0.485\,$\pm$\,0.026&44& 18562\,$\pm$\,754 & 0.305$_{-0.034}^{+0.040}$& 42\\
			
			J0842+3338 &19.3 &DA& 10779\,$\pm$\,182& 0.589\,$\pm$\,0.044&481& 10569\,$\pm$\,445 & 0.238$_{-0.158}^{+0.159}$ & 212\\
			
			J1637+3631 &10.0 &\tablenotemark{c} & & && 21209\,$\pm$\,1122 & 0.194$_{-0.283}^{+0.41}$& 18\\
			
			J0851+2713 &10.5 &DA& 13219\,$\pm$\,684& 0.757\,$\pm$\,0.088 &410 &16605\,$\pm$\,683 & 0.648$_{-0.209}^{+0.186}$& 155\\
			
			J1121+3200 &13.6 &DA& 13834\,$\pm$\,868&0.734\,$\pm$\,0.077 & 341 & 13489\,$\pm$\,925 & 0.839$_{-0.488}^{+0.278}$& 475\\
			
			J1133+4711 & 13.0&DA& 16713\,$\pm$\,1820&1.228\,$\pm$\,0.040 & 956 & 16545\,$\pm$\,1061 & 1.25$_{-0.055}^{+0.040}$& 1023\\
			
			J1138+5156 & 7.8&DA& 16588\,$\pm$\,928& 0.854\,$\pm$\,0.098 &276 &14538\,$\pm$\,918 & 1.055$_{-0.174}^{+0.112}$&706\\
			
			J1147+4758 & 9.5&DA& 15217\,$\pm$\,1024&0.684\,$\pm$\,0.100 & 228 & 17689\,$\pm$\,808 & 0.720$_{-0.368}^{+0.281}$&157\\
			
			J1157+4701 & 9.0&DA& 30584\,$\pm$\,898& 0.581\,$\pm$\,0.083  & 9 &26842\,$\pm$\,1437 & 0.511$_{-0.207}^{+0.278}$&13\\
			
			J1231+3312 & 15.1&DC& & & &18022\,$\pm$\,819 & 1.214$_{-0.053}^{+0.040}$&756\\
			
			J1238+3325 &15.0 &DA& 28128\,$\pm$\,774&0.628\,$\pm$\,0.065& 12 & 27663\,$\pm$\,1488 & 0.384$_{-0.214}^{+0.302}$ &13\\
			
			J1303+3338 &13.5 &DZ&10635\,$\pm$\,363 & 0.664$^{+0.197}_{-0.162}$ &  & 12517\,$\pm$\,640 & 0.829$_{-0.188}^{+0.145}$& 649\\
			
			J1345+4001 &13.6 &DA& 17818\,$\pm$\,436&0.569\,$\pm$\,0.043 & 107 & 18401\,$\pm$\,787 & 0.609$_{-0.072}^{+0.074}$& 93\\
			
			J1728-2329 &9.5 &DC: & & && 23573\,$\pm$\,1284 & 1.255$_{-0.0.69}^{+0.046}$&412\\
			
			J1430+2839 &8.9 &DA& 11050\,$\pm$\,327&0.657\,$\pm$\,0.095 & 525 & 10223\,$\pm$\,381 & 0.477$_{-0.206}^{+0.229}$&432\\
			
			J1400+2036 &20.1 &Unknown & & && 19890\,$\pm$\,1112 & 0.277$_{-0.169}^{+0.846}$&1011\\
			
			J2328+2138 &17.0 &DA& 18918\,$\pm$\,514&0.571\,$\pm$\,0.046 & 70 & 18284\,$\pm$\,833 & 0.891$_{-0.315}^{+0.196}$&229\\
			
		\end{tabular}
		\label{table:fits_gemini}
		\tablenotetext{a}{SNR reported as the average SNR per pixel between 4400 and 4600\,\AA}
		\tablenotetext{b}{Photometric values assume [He/H] = 0}
		\tablenotetext{c}{ See \cite{Raddi2019} for further details on classification}
	\end{table*}
	
	The resulting data were processed using the Gemini \textsc{PyRAF} package, which includes flat-fielding, bias correction, wavelength calibration, sky subtraction, and flux calibration using a spectral standard (Feige 66) which had been observed using an identical instrument setup. The spectra are then extracted and combined within the \textsc{PyRAF} environment using \textsc{SCombine}.

	The resulting spectra can be seen in Figure \ref{fig:spectra}, with the DAs (hydrogen lines present), and DZ (metal lines present) highlighted in black and blue respectively. Our sample contains 13 DA, 1 DZ, 2 DC (no features present), and 2 peculiar spectra which will be discussed in the following subsection. Also highlighted in Figure \ref{fig:spectra} are the gaps in the detector (grey shaded region) and the rest frame Balmer series from H$_{\alpha}$ to H$_{8}$ (dashed lines).
	
	\subsection{Peculiar Spectra}
	
	Our sample also contains two spectra with features inconsistent with being either a DA, DC, or DZ. These objects can be seen in Figure \ref{fig:selection} as green squares. The observational parameters for these objects (J1637+3631 and J1400+2036) can be found in Table \ref{table:observations}. Furthermore, their spectra can be seen in Figure \ref{fig:contamination}, along with the two DCs. A number of common white dwarf atmospheric lines are shown for clarity. 
	
	One of our objects, J1637+3631 was studied in detail in \cite{Raddi2019}. This object shows evidence of oxygen, magnesium, silicon, and calcium absorption, which are peculiar elements to find within a white dwarf atmosphere. The authors suggest that this star is a white dwarf, likely a subtype DS (oxygen features present, see \cite{Williams2019}) given the abundance of oxygen in its atmosphere, the second of its kind after J1240+6170 \citep{Kepler2016}. This white dwarf may represent the remnant of an intermediate-mass star which forms an oxygen-neon white dwarf as opposed to the typical carbon-oxygen. Similar spectral features are also seen in J1400+2036, suggesting that it could also belong to this class of remnants. These objects both show large tangential velocities (300-500 kms$^{-1}$) that allowed them to pass our selection criteria.
	
	Given the relative scarcity of these objects, and the difficulty in modeling their atmospheric composition, we do not attempt to determine the mass or age of these objects. Therefore, we continue our analysis using the DA, DC, and DZ objects unless otherwise noted.

	\begin{figure*}[!t]
		\includegraphics[angle=0,width=\textwidth]{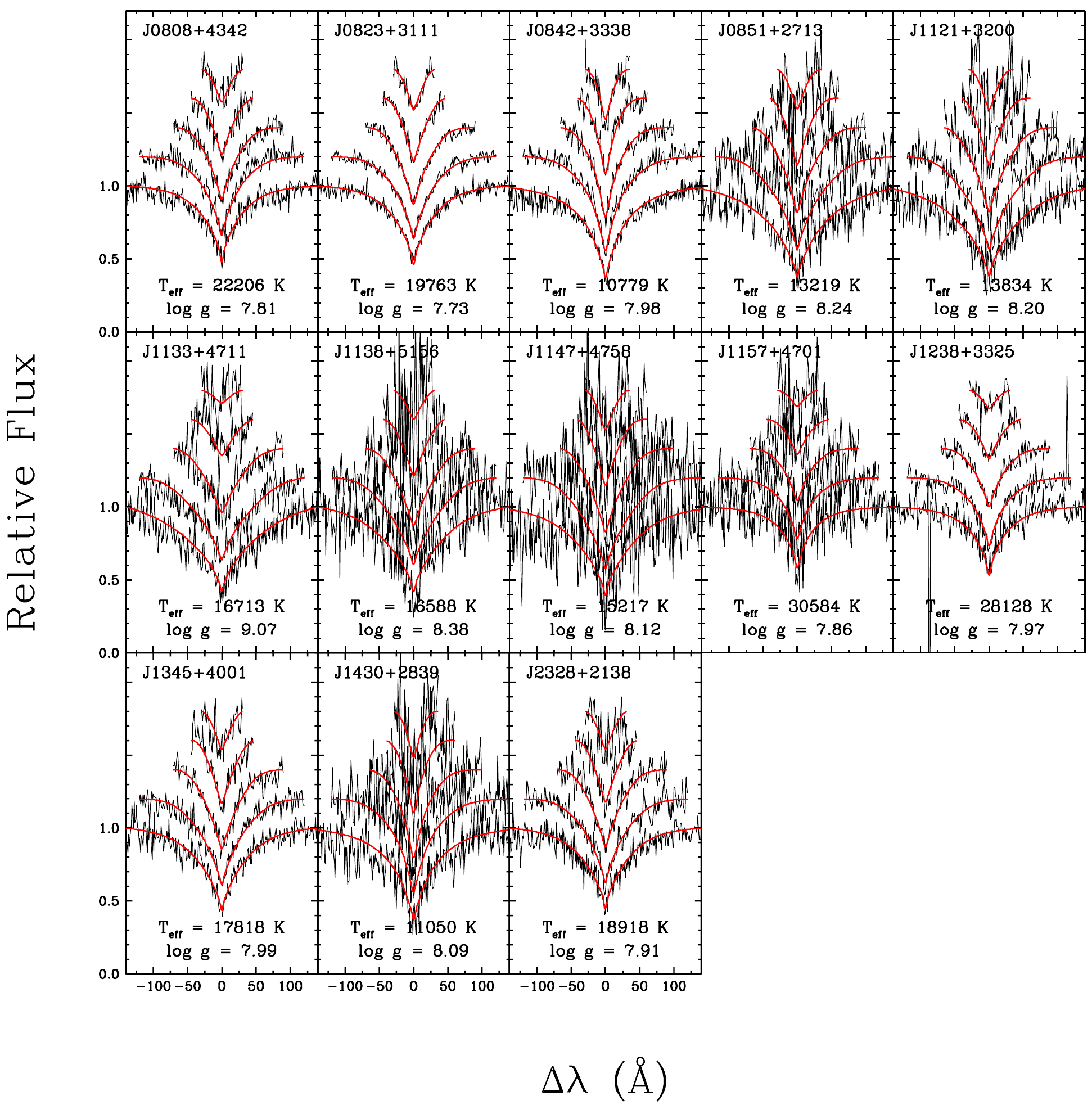}
		
		\caption{Fit to the spectra obtained from GMOS for the DA sample. Each panel shows the normalized Balmer lines, from H$_{\beta}$ to H$_8$ (black) with the fit over-plotted (red). Temperature and surface gravity values are 3D corrected using the method presented in \cite{Tremblay2013}. }
		
		\label{fig:fit}
	\end{figure*}
	
	\begin{figure*}[!t]
		\includegraphics[angle=0,width=0.49\textwidth]{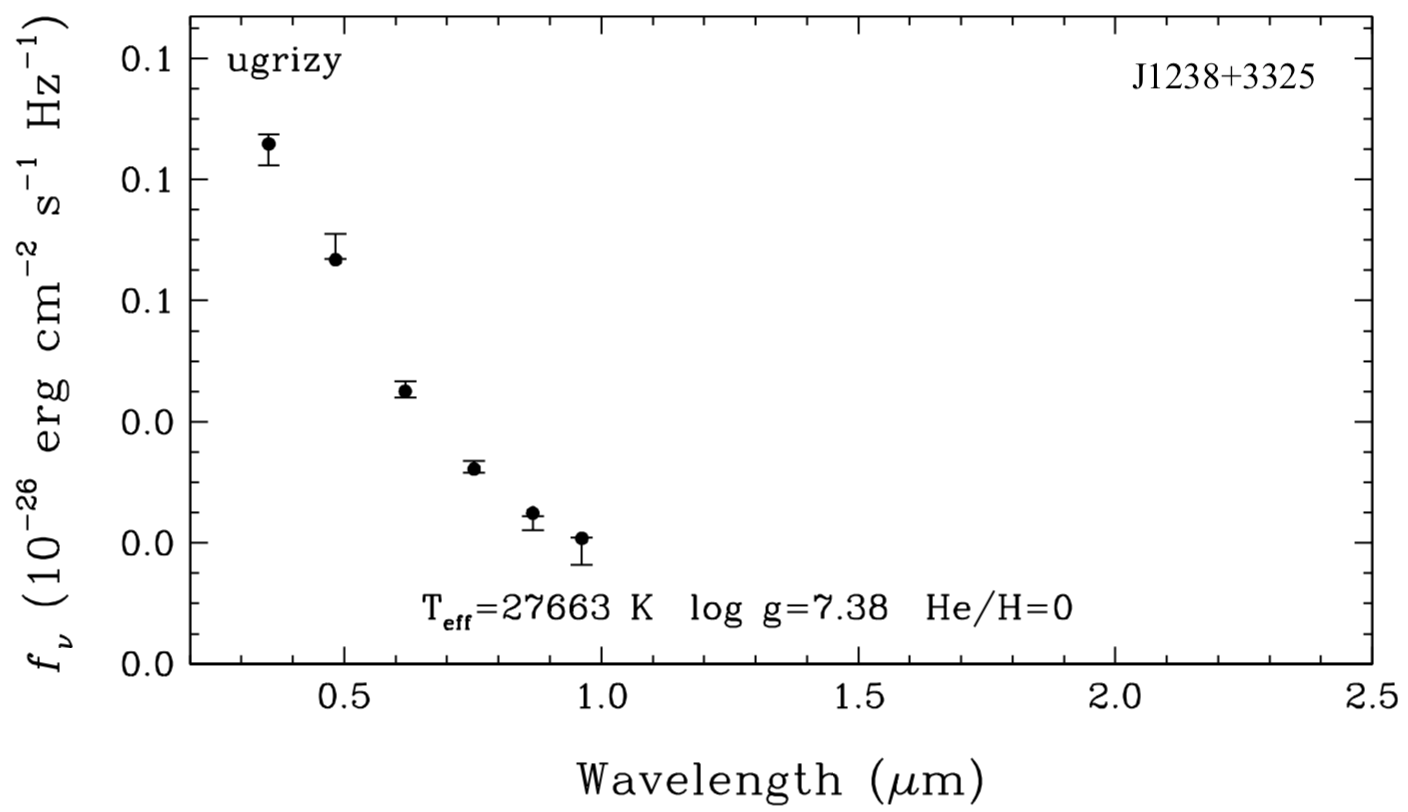}
		\includegraphics[angle=0,width=0.49\textwidth]{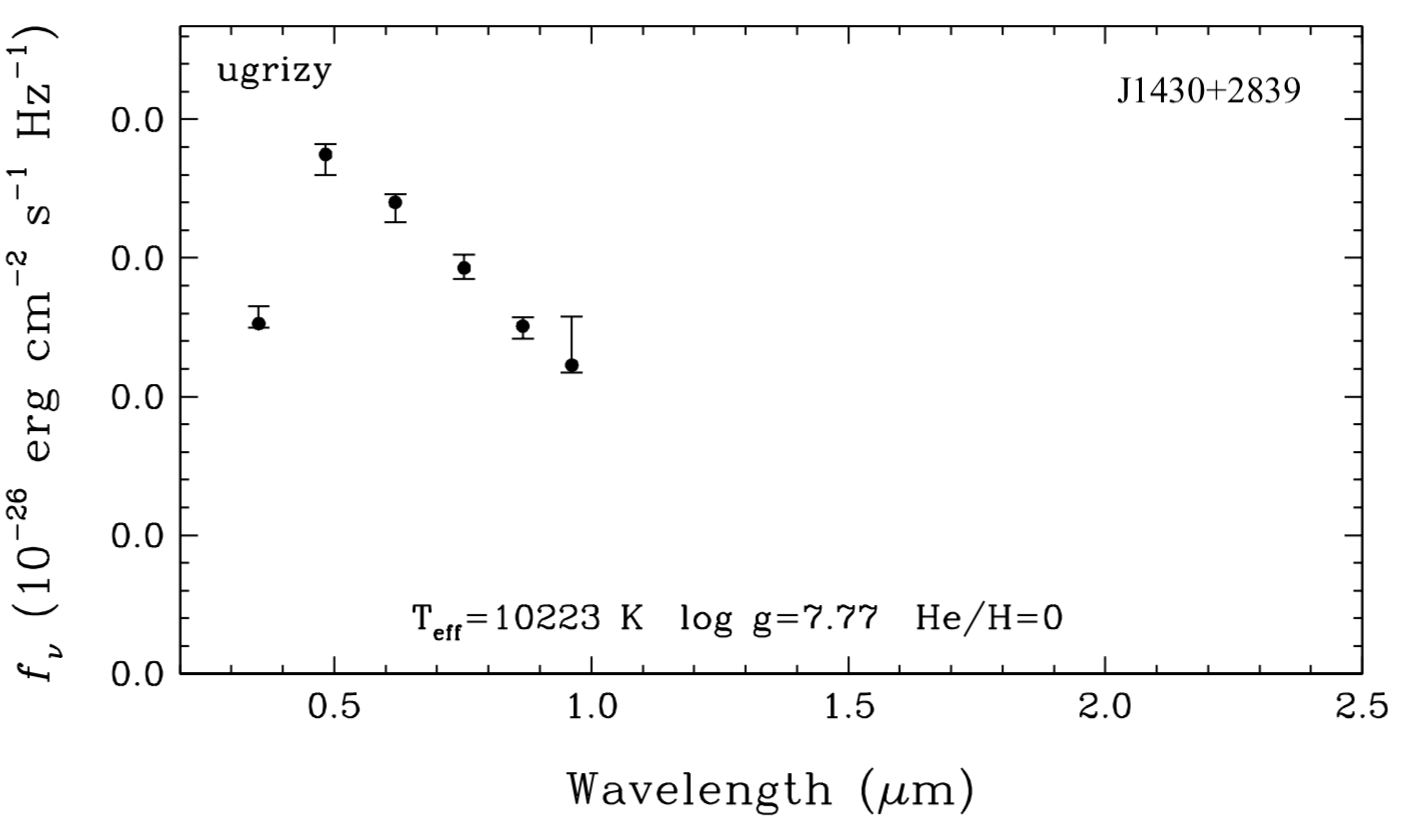}
		
		\caption{An example of the photometric technique used to measure the mass and temperature using the photometry in combination with the \textit{Gaia} parallax. The photometry used includes the CFIS $u$ in combination with the PS1 $grizy$ bands. The observed magnitudes are represented by the error bars, and the photometry of the best fit model is indicated by the circles. The resulting best fit model parameters are also presented. 
		}
		\label{fig:photometric_fit}
	\end{figure*}
	
	\begin{figure*}[!t]
		\includegraphics[angle=0,width=\textwidth]{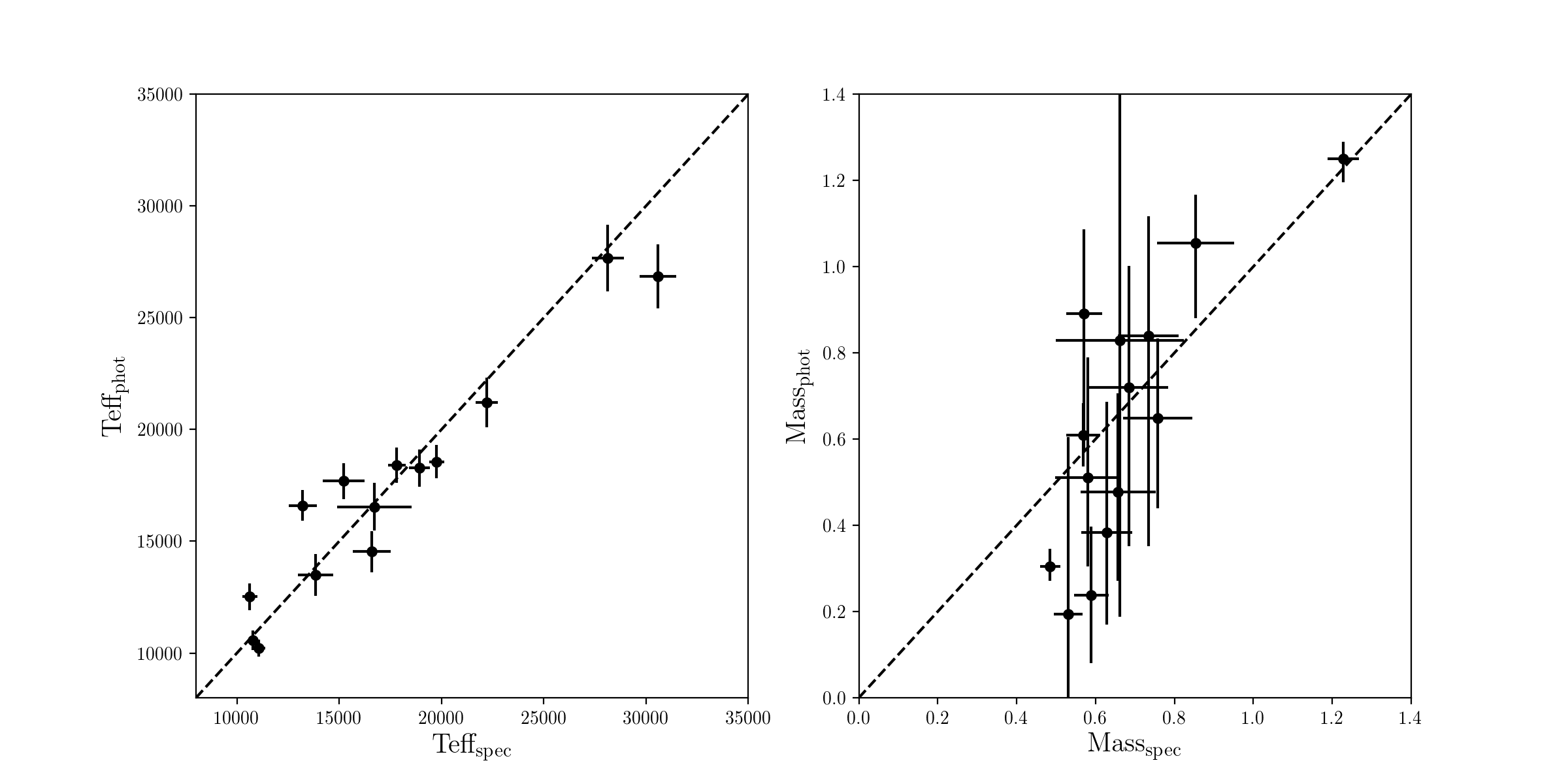}
		\caption{Comparing the spectroscopic and photometric temperature (left) and mass (right) determinations for the 14 DA and DZ white dwarfs with spectroscopic masses.}
		\label{fig:Parameter_comparison}
	\end{figure*}

	\section{Temperature and Mass Estimation}
	\label{sec:Results}
	
	\subsubsection{DA}
	We obtain temperatures (T$_{\textrm{eff}}$) and surface gravities (log $g$) by fitting the observed Balmer lines with model atmospheres using the technique originally presented in \cite{Bergeron1992}. The temperature and surface gravity values have been corrected for 3D effects using the prescription from \cite{Tremblay2013}. The surface gravity is then converted to a mass using the white dwarf mass-radius relationship. This technique has been used in a number of previous studies to measure the masses of DA white dwarfs \citep[see, e.g,][]{Tremblay2011, Genest2019}. The resulting fits can be seen in Figure~\ref{fig:fit}. 
	
	The reported errors on the white dwarf parameters take both internal and external errors associated with the fitting technique. The internal errors are based on how well the model fits the data, with this error approaching zero as the SNR approaches infinity. The external errors are estimated to be 1.2\% in T$_{\textrm{eff}}$ and 0.038 dex in log $g$ as determined by \cite{Liebert2005}. The total error is then the sum in quadrature of the internal and external errors.
	
	We apply this technique to our sample of 13 DA white dwarfs and present the results in Table~\ref{table:fits_gemini}. 
	
	\subsubsection{DZ}
	
	Analysis of the DZ (J1303+3338) was performed in the same manner as described in \cite{Coutu2019} based on updated methods from \cite{Dufour2005,Dufour2007}. This method uses a combination of the photometric method (photometry and parallax) and the spectrum to determine the calcium abundance. The resulting fit also returns the temperature and mass, which are presented in Table \ref{table:fits_gemini}. The resulting mass is quite uncertain due to the poor parallax measurement. As a result, we proceed with only the DA sample.

	\begin{figure*}[!t]
		\includegraphics[angle=0,width=\textwidth]{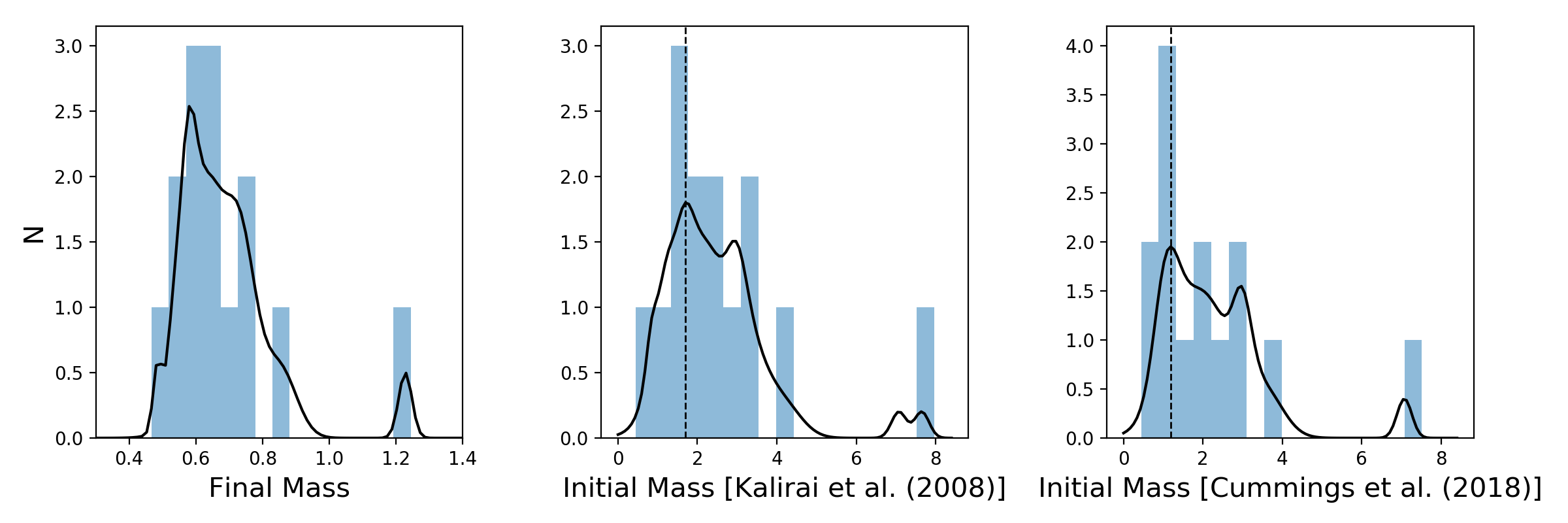}
		
		\caption{\textit{Left:} The distribution of white dwarf masses as displayed in Table \ref{table:fits_gemini}. \textit{Middle:} Initial masses calculated using the \cite{Kalirai2008} IFMR for our sample of white dwarfs. The mean initial mass for the sample is indicated by a dashed line. \textit{Right:} Same as the middle panel except the IFMR of \cite{Cummings2018} was used.}
		\label{fig:initial_mass}
	\end{figure*}

	\subsection{Photometric Technique}
	
	Another technique that can be used to determine the surface temperature and mass of a white dwarf relies on the distance and spectral energy distribution. This technique, called the photometric method, uses the theoretical magnitudes from model atmospheres and compares them to broadband photometric observations \citep{Bergeron1997}. For this method we use the CFIS $u$ in combination with the PS1 $grizy$ bands. These magnitudes were de-reddened using E(B - V) values from \cite{Schlegel1998} combined with extinction coefficient values from \cite{Schlafly2011} assuming R$_{\mathrm{V}}$ = 3.1. For the CFIS $u$-band the extinction coefficient was determined to be 4.14. The $u$-band in particular is important for determining the temperatures of hot white dwarfs \citep{Genest2014} as they have a considerable amount of ultraviolet flux \citep{Bergeron2019}.
	
	The best fit returns the temperature and the solid angle, $\pi(R/D)^2$. Therefore, the parallaxes from \textit{Gaia} DR2 can be used to determine the distance, allowing one to solve for the radius. The mass is then determined using the mass-radius relation as in the spectroscopic technique. We present these results in the final two columns of Table \ref{table:fits_gemini}.
	
	An example of the fitting results can be seen in Figure \ref{fig:photometric_fit}, which shows the photometry as error bars and the resulting synthetic photometry as black circles. The left-hand panel shows the results for a hot white dwarf (J1238+3325), and the right-hand panel shows a fit for a cooler white dwarf (J1430+2839). The resulting temperature of both objects agrees with the spectroscopic values, and reiterate the importance of the $u$-band as the photometric temperature is determined to be nearly 10,000 K hotter for J1238+3325 when it is neglected.
	
	The resulting photometric temperatures and masses are then compared to their spectroscopic counterparts in Figure \ref{fig:Parameter_comparison}. The temperatures agree quite well between both methods, however, the masses show much larger scatter given the large uncertainties in the \textit{Gaia} parallax values for some of the objects. The photometric technique does correctly pick out the most massive white dwarf in our sample (J1133+4711), but many of the remaining objects have essentially unconstrained masses due to poor parallax measurements. For this reason, the remainder of this work will focus on the resulting masses obtained spectroscopically unless otherwise noted.

	\subsection{Mass Distribution}
	\label{subsec:Masses}

	The resulting spectroscopic and photometric masses are presented in Table \ref{table:fits_gemini}. In the left-hand panel of Figure \ref{fig:initial_mass} we show the distribution of white dwarf masses by representing each measurement as a Gaussian with a mean equal to the measured mass and the standard deviation equal to the corresponding uncertainty. The distribution peaks at 0.58 M$_{\odot}$ and contains a few higher and low mass objects, including J1133+4711 with a mass of 1.22\,$\pm$\,0.04\,M$_{\odot}$, and J0823+3111 with a mass of 0.49\,$\pm$\,0.03\,M$_{\odot}$.
	
	This study focuses on the young, hot, white dwarfs that most recently formed in the halo. This means that under the typical assumption that the halo formed $\sim$12\,Gyr ago during a short burst of star formation, one would expect these young white dwarfs to have masses of $0.50 - 0.58$\,M$_{\odot}$ \citep{Kalirai2008}, and hence such a large spread was not expected.

	\subsection{Ages}
	\label{subsec:Ages}
	
	\subsubsection{Ages using Stellar Isochrones}
	
	Surprisingly, many of the objects do not have masses consistent with those of white dwarfs in the oldest globular clusters, suggesting that they do not have ages consistent with what is normally believed to be an old ($\sim$12\,Gyr) halo population.
	
	In order to calculate the total ages we require a progenitor lifetime to add to the calculated white dwarf cooling times presented in Table \ref{table:fits_gemini}. We begin by calculating the progenitor mass using the IFMR. We choose to compare two IFMRs, those from \cite{Kalirai2008} and \cite{Cummings2018}. The main difference between the two is that for a given initial mass the \cite{Cummings2018} IFMR returns systematically higher white dwarf masses on the order of 0.05\,M$_{\odot}$. Equivalently, this means that for a given white dwarf mass the \cite{Cummings2018} IFMR will return lower initial masses, and hence longer pre-white dwarf ages. The resulting initial masses for both IFMRs can be seen in the middle and right-hand panels of Figure \ref{fig:initial_mass}.
	
	With the initial mass calculated, the remaining step is to calculate the progenitor's lifetime. This requires an assumption regarding the metallicity combined with a stellar isochrone. For this study, we use two isochrones: the PAdova and tRieste Stellar Evolution Code (PARSEC) \citep{PARSEC} and the Mesa Isochrones and Stellar Tracks (MIST) \citep{Choi2016}. We adopt an initial metallicity of [Fe/H] = $-$1.5 dex for our progenitors, which is typical for a halo population \citep{Peng2013}.
	
\begin{figure*}[!t]
		\includegraphics[angle=0,width=\textwidth]{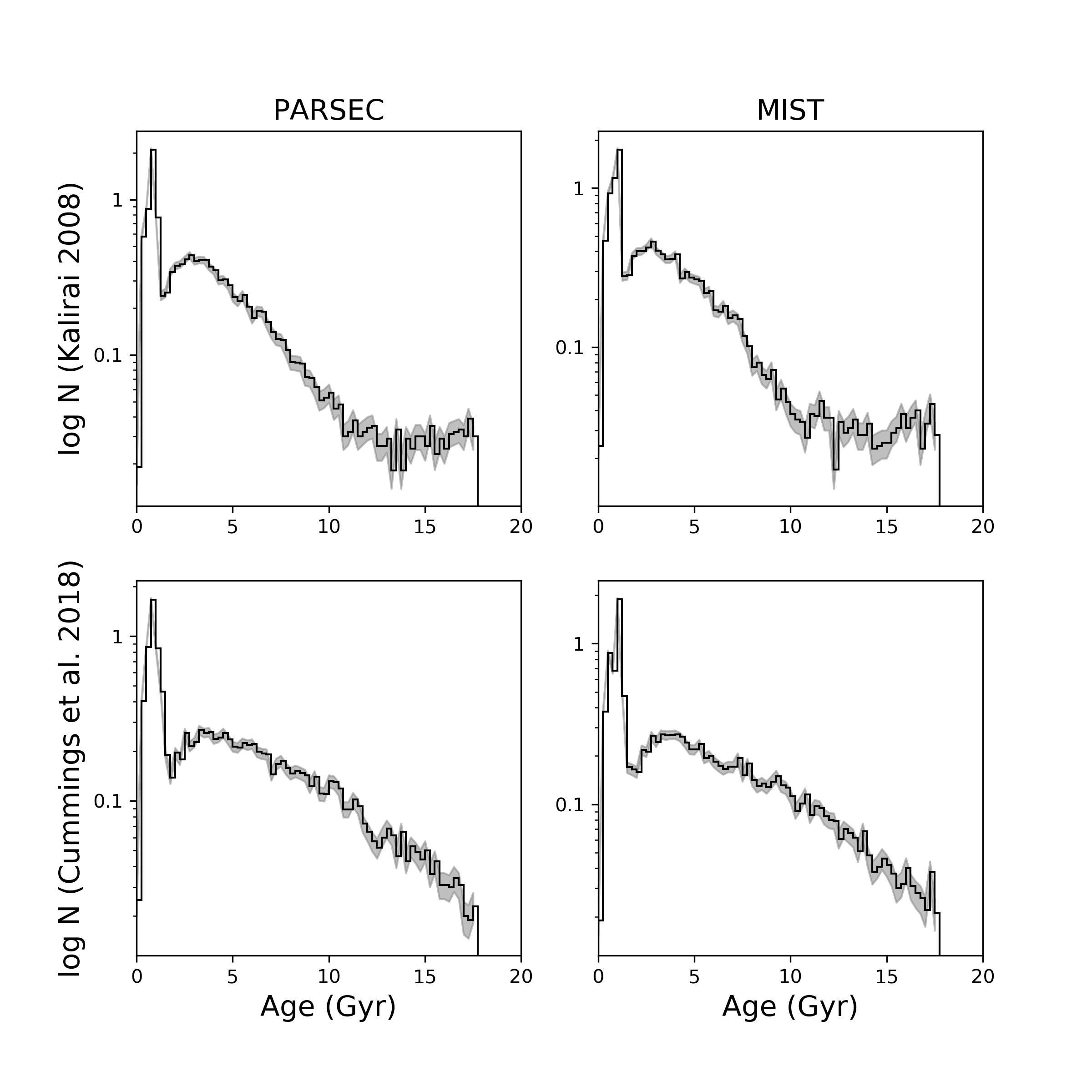}
		
		\caption{Age distribution for our 14 white dwarfs with spectroscopically confirmed masses. Each object is plotted as a Gaussian with a mean equal to the age presented in Table \ref{table:ages_gemini} and standard deviation equal to the error.}
		\label{fig:ages}
	\end{figure*}

	With a choice of metallicity and isochrone, we calculate the progenitor age and add it to the white dwarf cooling age to obtain the total age of each star. We then propagate the uncertainties in the white dwarf mass and cooling ages through the IFMR and stellar isochrone to obtain the distribution of total ages and present the sum of these histograms in Figure \ref{fig:ages}. We note that the choice of IFMR has a much larger effect on the total age relative to the choice of isochrone, and as such we will only present results with the MIST isochrones. Each object is plotted as a Gaussian with the mean equal to the age, and the standard deviation equal to the error in the age. We also present the resulting ages using the various combination of isochrones and IFMRs in Table \ref{table:ages_gemini}.
	
	\begin{table*}[!t]
		
		\centering
		\caption{Ages for Halo White Dwarfs  \\ \smallskip}
		
		\resizebox{\textwidth}{!}{\begin{tabular}{l|ccc|cccc}
				&    &   & &Age (Gyr) & Age (Gyr)&  Age (Gyr)& Age (Gyr)   \\
				
				ID  & T$_{\textrm{eff}}$(K)   & Mass  (M$_{\odot}$)& t$_{cool}$ (Myr)& MIST + K08 & PARSEC + K08 &  MIST + C18 & PARSEC + C18   \\
				\hline
				\smallskip
				
				J0808+4342 & 22206\,$\pm$\,543&0.531\,$\pm$\,0.036& 28 &10.3$\pm$ 3.7 & 10.1$\pm$ 3.6  & 12.5$\pm$ 4.5 & 12.8$\pm$ 4.1\\
				
				J0823+3111 & 19763\,$\pm$\,366& 0.485\,$\pm$\,0.026& 44 &18.9$\pm$ 2.7 & 18.5$\pm$ 2.7  & 14.6$\pm$ 2.2  &14.8$\pm$ 2.3\\
				
				J0842+3338 & 10779\,$\pm$\,182& 0.589\,$\pm$\,0.044&481 &3.8$\pm$ 1.2 &  3.7$\pm$ 1.1 & 6.5$\pm$ 2.5  &6.7$\pm$ 2.5\\
				
				J0851+2713 & 13219\,$\pm$\,684& 0.757\,$\pm$\,0.088 &410 &0.9$\pm$ 0.2 & 1.0$\pm$ 0.2  & 1.1$\pm$ 0.2  &1.1$\pm$ 2.3\\
				
				J1121+3200 & 13834\,$\pm$\,868&0.734\,$\pm$\,0.077 & 341 & 0.9$\pm$ 0.2& 0.9$\pm$ 0.2   & 1.1$\pm$ 0.2  &1.1$\pm$ 0.2\\
				
				J1133+4711 & 16713\,$\pm$\,1820&1.228\,$\pm$\,0.040 & 956 & 1.0$\pm$ 0.0 & 1.0$\pm$ 0.0   &  1.0$\pm$ 0.0 &1.0$\pm$ 0.0\\
				
				J1138+5156 &16588\,$\pm$\,928& 0.854\,$\pm$\,0.098 &276 &  0.5$\pm$ 0.1&  0.5$\pm$ 0.1  &  0.5$\pm$ 0.1 &0.5$\pm$ 0.0\\
				
				J1147+4758 &15217\,$\pm$\,1024&0.684\,$\pm$\,0.100 & 228 & 2.1$\pm$ 0.8 &  2.1$\pm$ 0.7  & 2.9$\pm$ 1.2  &2.9$\pm$ 1.2\\
				
				J1157+4701 &30584\,$\pm$\,898& 0.581\,$\pm$\,0.083  & 9 & 6.2$\pm$ 2.7&  6.2$\pm$ 2.7  & 8.5$\pm$ 3.7  &8.2$\pm$ 3.8\\
				
				J1238+3325 & 28128\,$\pm$\,774&0.628\,$\pm$\,0.065& 12&2.4$\pm$ 1.0  & 2.3$\pm$ 1.0  & 3.9$\pm$ 1.7  &3.9$\pm$ 1.7 \\
				
				J1303+3338 &10635\,$\pm$\,363 & 0.664$^{+0.197}_{-0.162}$ & 649&5.2$\pm$ 2.2 & 5.1$\pm$ 2.1   & 5.6$\pm$ 2.5  & 6.1$\pm$ 2.6\\
				
				J1345+4001 & 17818\,$\pm$\,436&0.569\,$\pm$\,0.043 & 107&5.2$\pm$ 1.8 & 5.1$\pm$ 1.8   &  9.1$\pm$ 3.6 &9.0$\pm$ 3.7\\
				
				J1430+2839 &11050\,$\pm$\,327&0.657\,$\pm$\,0.095 & 525 &  3.1$\pm$ 1.1 &  3.0$\pm$ 1.1  &  4.1$\pm$ 1.6 &4.1$\pm$ 1.7\\
				
				J2328+2138 & 18918\,$\pm$\,514&0.571\,$\pm$\,0.046 & 70 & 5.1$\pm$ 1.9& 5.1$\pm$ 1.8   & 8.6$\pm$ 3.5  &8.8$\pm$ 3.6\\
				
			\end{tabular}}
			\label{table:ages_gemini}
		\end{table*}

		Our results show that the majority of our sample have ages inconsistent with other estimates of the age of the Galactic halo, irrespective of our choice of isochrone or IFMR. In particular, many of the total ages of our objects are on the order of $1 - 2$ Gyr. Only six having ages consistent with being $>$ 10 Gyr, although one (J0823+3111) has an age much larger than the age of the Universe.

		\subsubsection{Ages using Kalirai (2012)}
		
		The IFMR is typically calculated empirically using star clusters as the constant age of the stars means that the initial and final masses can both be accurately measured. Observationally, however, calibrating the IFMR at low initial mass is difficult as the old clusters are typically located at large distances from the Sun, making even the hot white dwarfs in the cluster very faint. This led \cite{Kalirai2012} to construct a relationship between the final white dwarf mass using the masses of four field halo white dwarfs and the mean mass of six white dwarfs from M4,
		
		\begin{equation}
			\label{eq:kalirai2012}
			\log[\textrm{Age (Gyr)}] = (\log[M_{\textrm{final}}/M_{\odot} + 0.270] - 0.201)/0.272.
		\end{equation}
		
		Applying this relation to field stars comes with certain caveats. Most significantly, this relation is only calibrated for a specific mass and metallicity range which may not be true for field halo stars.

\begin{figure}[!t]
			\includegraphics[angle=0,width=0.5\textwidth]{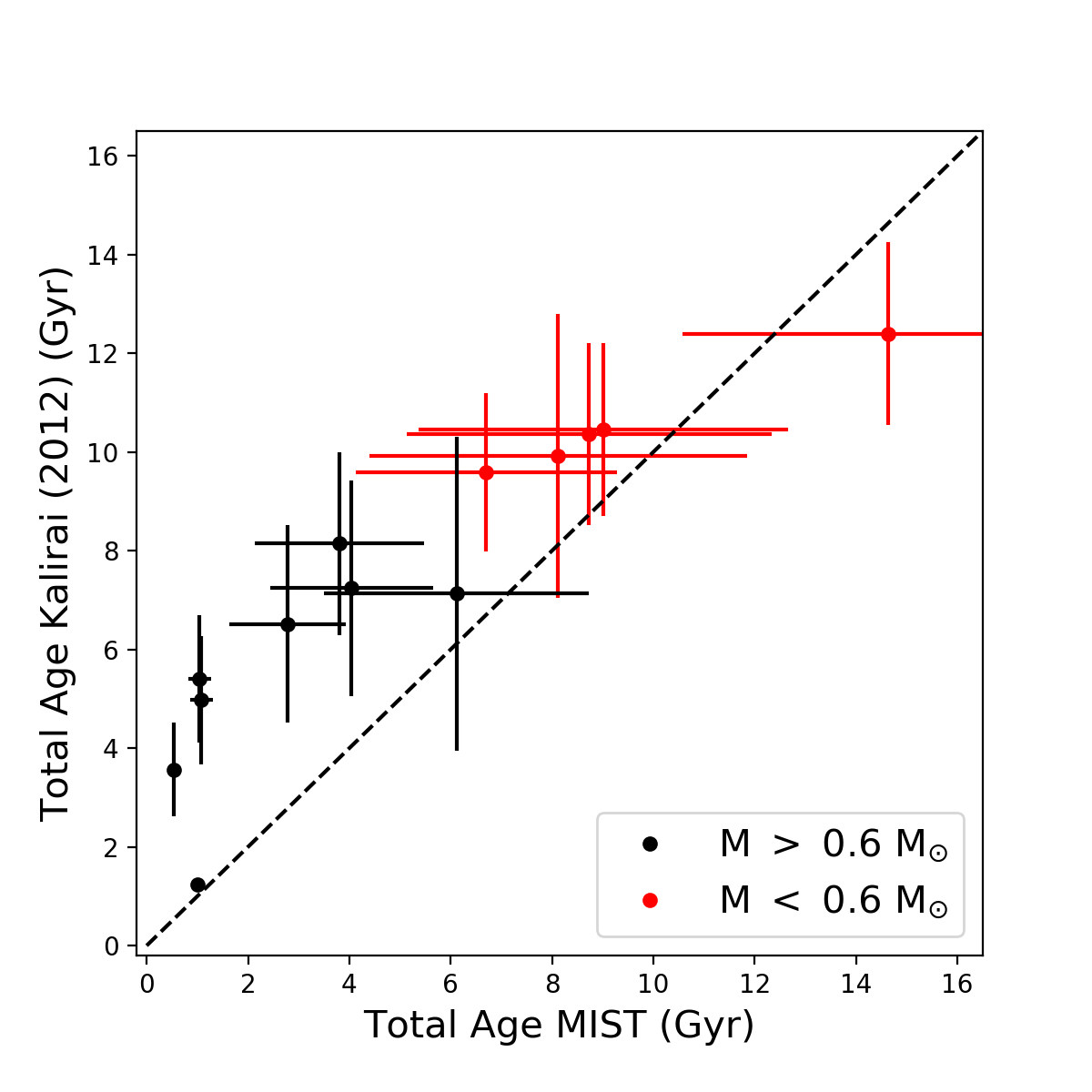}
			
			\caption{Total ages using Equation \ref{eq:kalirai2012}, with objects of mass greater (black) less than (red) 0.6\,M$_{\odot}$ highlighted compared to the ages obtained using the \cite{Cummings2018} IFMR and MIST isochrones.}
			\label{fig:kalirai_ages}
		\end{figure}

		We apply this relation to our sample, noting that many of our objects do not fall within this calibrated relation, and therefore incur a systematic bias. Indeed, Figure \ref{fig:kalirai_ages} shows that the ages of our higher mass objects increase significantly, essentially forcing them into an age compatible with the traditional view of a 12-14\,Gyr halo. Therefore, we continue with the ages derived using the IFMR.

		\subsection{Uncertainties}
		\label{sec:bias}
		
		As Table \ref{table:fits_gemini} shows, a number of our objects have mass uncertainties of 0.05-0.1\,M$_{\odot}$. At the high-mass end, this corresponds to a fairly small uncertainty on the age (i.e., the difference in main-sequence age between a 5\,M$_{\odot}$ and 6\,M$_{\odot}$ star is only $\sim$30\,Myr) whereas, at low initial mass, this can lead to uncertainties on the order of the age of the Universe. Take, for example, J1147+4758, which has a mass of 0.684\,$\pm$\,0.1\,M$_{\odot}$. Using the \cite{Cummings2018} IFMR, this object descended from a star with an initial mass of $1.1 - 3.2$\,M$_{\odot}$, which have main-sequence lifetimes of $0.3 - 4.5$\,Gyr.

\begin{figure*}[!t]
			
			\includegraphics[angle=0,width=\textwidth]{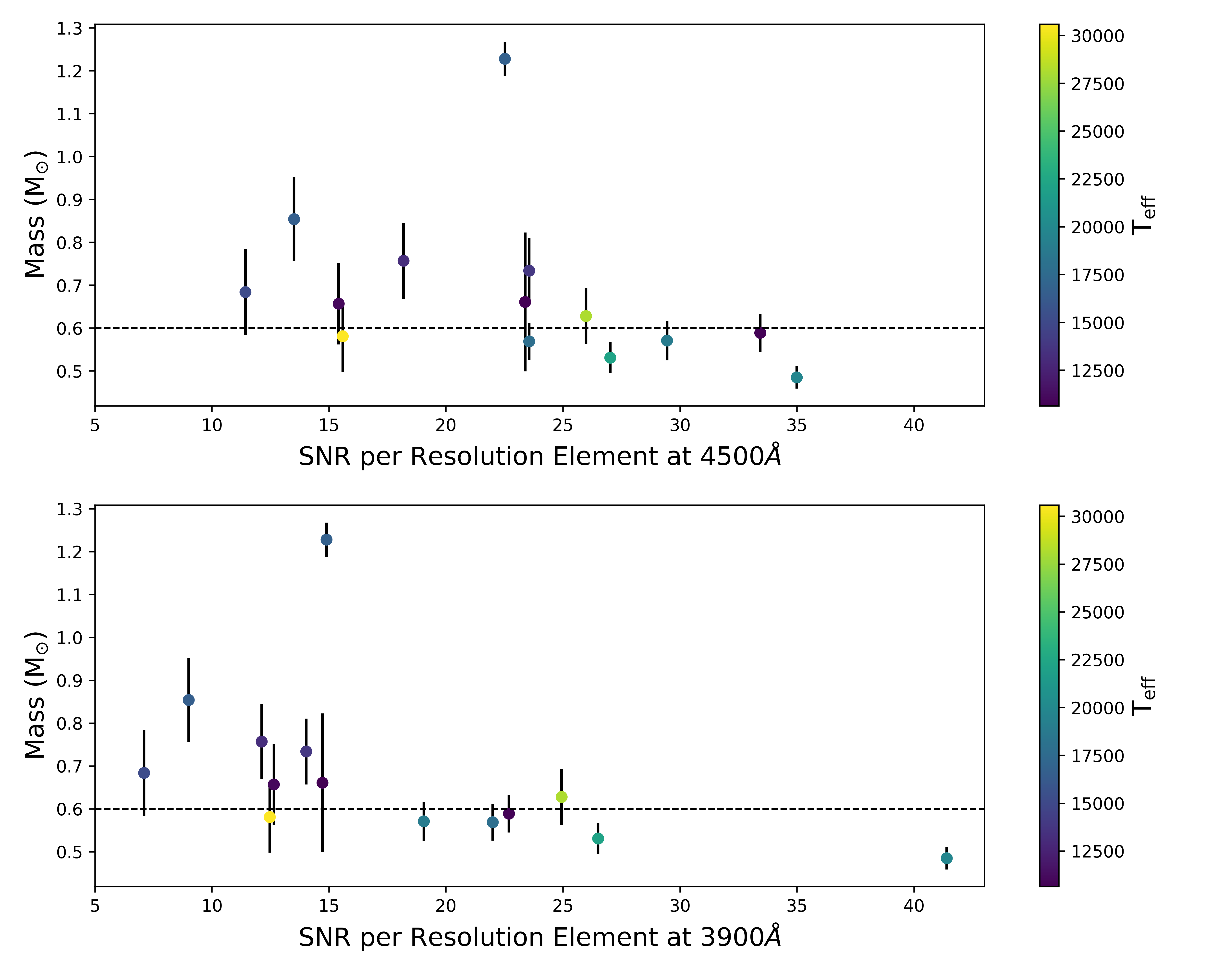}
			\caption{White dwarf mass vs. signal-to-noise (SNR), colour-coded by temperature. The top panel shows the SNR calculated at H$_{\beta}$, while the bottom plot shows the SNR calculated at H$_8$. The lower plot, in particular, shows evidence for a systematic decline in derived mass with increasing SNR.}
			\label{fig:snr}
		\end{figure*}

		Interestingly, it seems that many of the objects with large uncertainties also have masses that are higher than expected for an old halo population. In Figure \ref{fig:snr}, we show this effect by plotting the SNR per resolution element against derived white dwarf mass. The SNR here is calculated in two places: (1) at H$_{\beta}$ (top) and (2) at H$_8$ (bottom). Under the assumption that these objects are truly part of the halo (see Section \ref{sec:halo_discussion}), both plots suggest an underestimation in the mass uncertainty in the fitting routine as higher SNR values, particularly at H$_8$, have lower white dwarf masses. Figure \ref{fig:snr} also shows that all of the high-mass objects do not pass the criteria we set out to achieve for the observations as they have SNR values below 15 at H$_8$.

\begin{figure*}[!t]
			
			\includegraphics[angle=0,width=\textwidth]{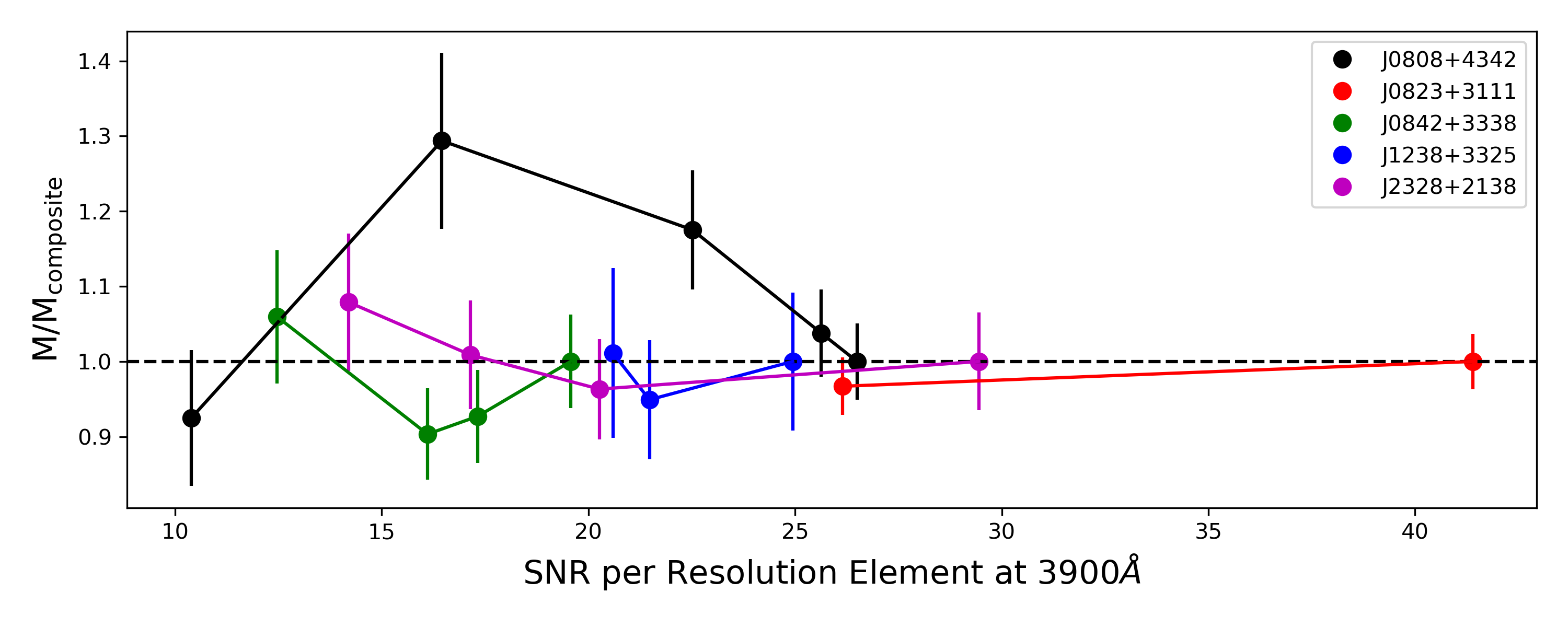}
			\caption{Mass determinations, as a fraction of the composite mass presented in Table \ref{table:fits_gemini}, for low mass sub-exposures. }
			\label{fig:degraded}
		\end{figure*}

		\cite{Kepler2006} used Gemini spectra of white dwarfs near 12,000\,K, the location of the maximum Balmer line equivalent width, and showed that for a given object, higher SNR data at H$_8$ returned lower white dwarf masses. We attempt to test this by fitting combinations of sub-exposures for objects with high SNR and low mass to see if their mass decreases with increased SNR. The results are shown in Figure \ref{fig:degraded}, where, for each object, one sub-exposure has been added per increase in SNR. The results are inconclusive, however, as some objects show an increase in mass as SNR increases while others show a decrease. Many of the sub-exposures are relatively high SNR observations, so only J0808+4342, J0842+3338, J2328+2138 had single exposures below our SNR threshold of 15. These objects show a 10\% decrease, a 7\% increase, and a 9\% increase, respectively. The explanation may lie in the flux calibration and this potential issue requires more careful considerations with larger sample sizes that have low SNR data.

		\section{Analyzing our Halo Sample}
		\label{sec:halo_discussion}
		\subsection{Halo Membership}
		\label{sec:velocities}
		
		\begin{figure*}[!t]
			
			\includegraphics[angle=0,width=\textwidth]{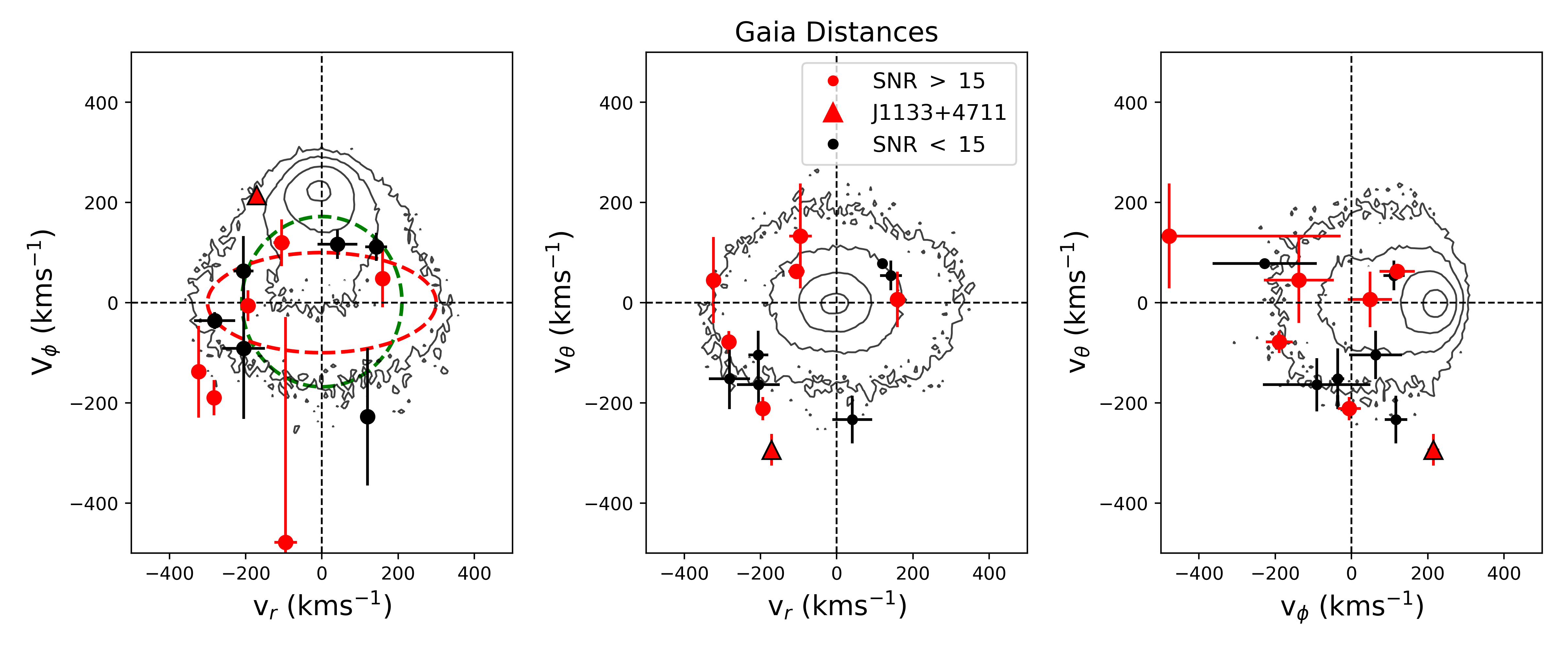}
			\includegraphics[angle=0,width=\textwidth]{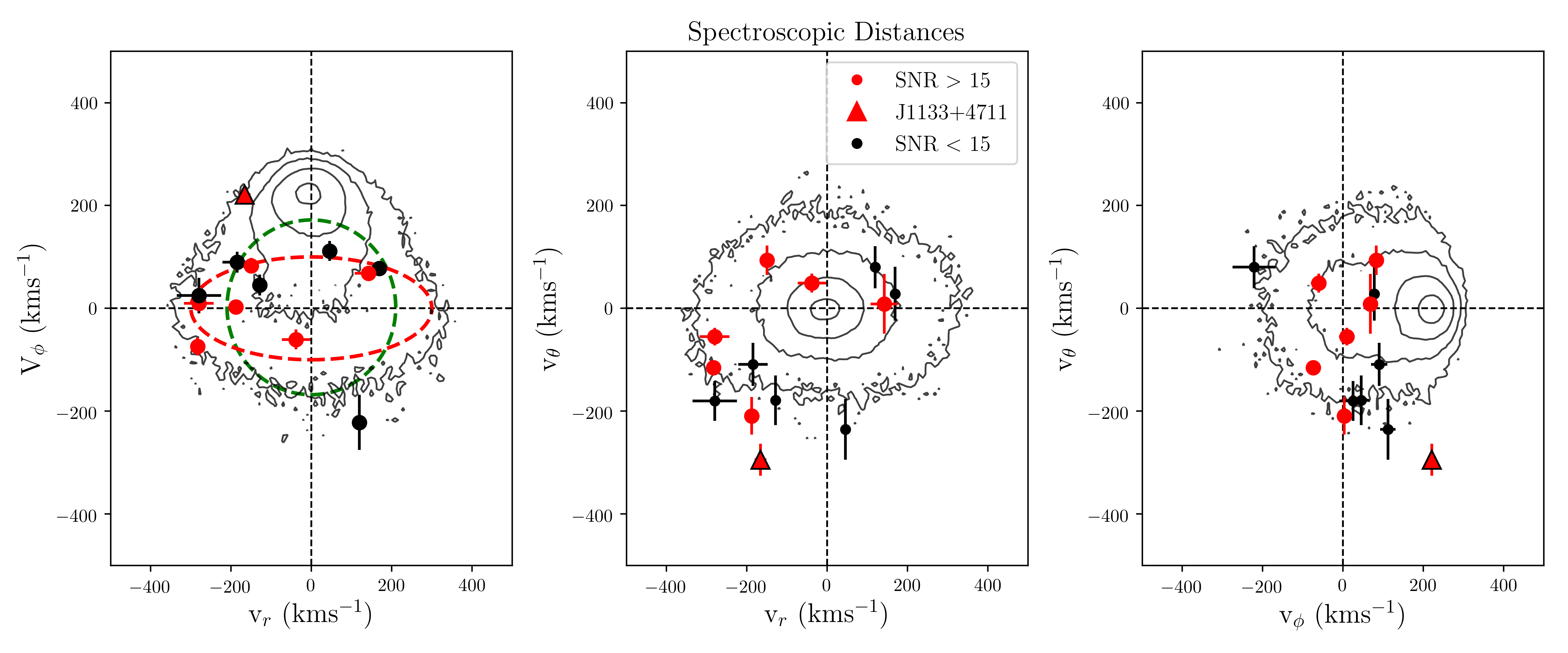}
			\caption{Spherical velocity distribution using distances calculated using parallax measurements (top) and the spectroscopically determined absolute magnitude (bottom). We also show a classical halo (green circle) and a more flattened halo (red oval) which is attributed to the Gaia-Enceladus merger event \citep{Fattahi2019}.}
			\label{fig:velocities}
		\end{figure*}
		
		Given that half of our sample have masses consistent with a thin disk origin, it is useful to first check whether their kinematics could be consistent with such a population. We use the dwarf/giant catalogue presented in \cite{Thomas2019} combined with radial velocities from LAMOST \citep{LAMOST, LAMOST2} as a reference catalog for other disk and halo stars. We use this information to calculate the 3D velocities in spherical coordinates and present them in Figure \ref{fig:velocities} as the black contours. The figure shows the disk centered at (v$_r$, v$_{\phi}$) = (0, 200), with the extended contours centered at (0,0) representing the halo population. Specifically, the green circle represents a classical isotropic population and the red dashed oval represents a radially-biased component recently dubbed the ``Gaia-Enceladus" \citep{Helmi2018,Belokurov2018, Fattahi2019}.
		
		Over-plotted on Figure \ref{fig:velocities} are our white dwarfs with distances calculated using the \textit{Gaia} parallaxes (top) and the spectroscopic distances (bottom). These distances are calculated using the absolute magnitude calculated as part of the spectroscopic fit combined with extinction calculated from the \cite{Schlegel1998} dust maps. Figure \ref{fig:velocities} shows that the objects with SNR $>$ 15 at H$_8$ do not populate a unique location within these diagrams, suggesting that both samples are on similar orbits. This means that the white dwarfs, with a possible exception of our high mass WD J1133+4711 (red triangle), have velocities inconsistent with a disk origin as they are on highly radial orbits with minimal rotation.
		
\begin{figure*}[!t]
			
			\includegraphics[angle=0,width=\textwidth]{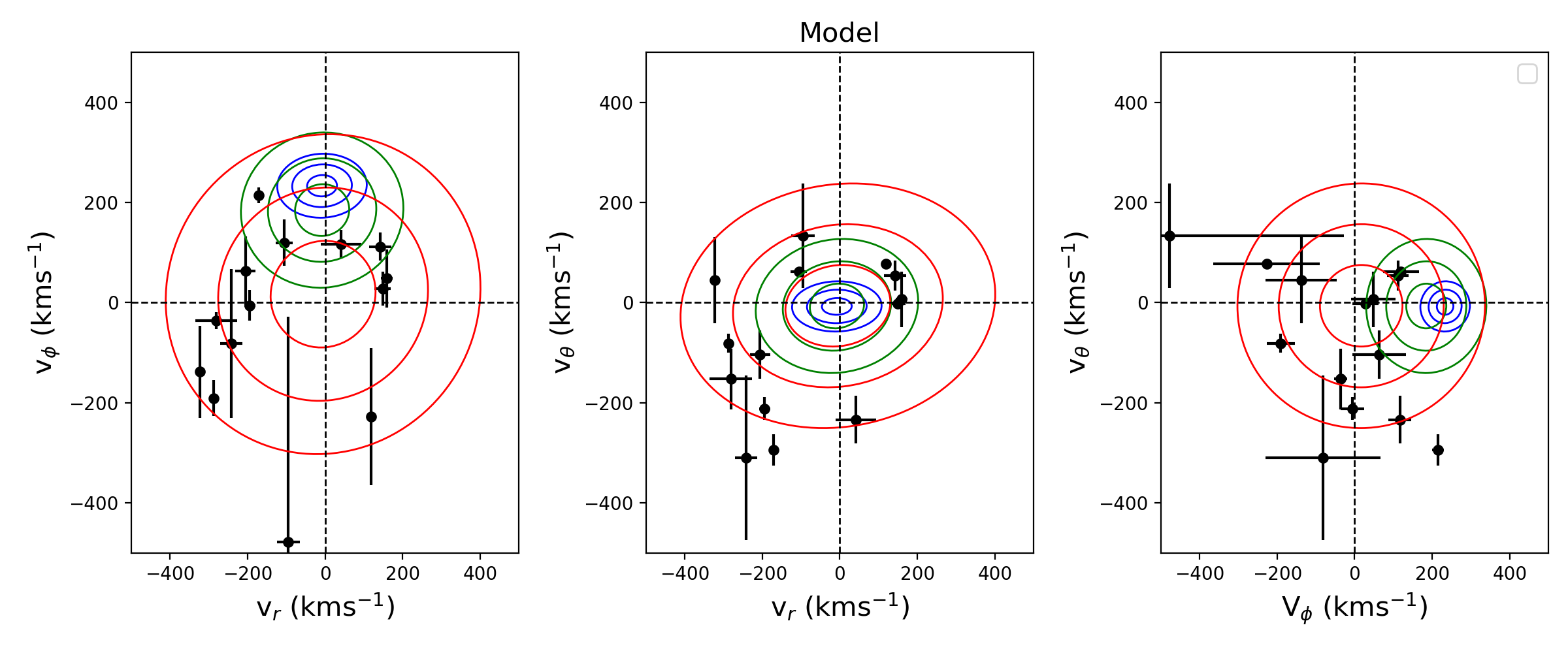}
			\includegraphics[angle=0,width=\textwidth]{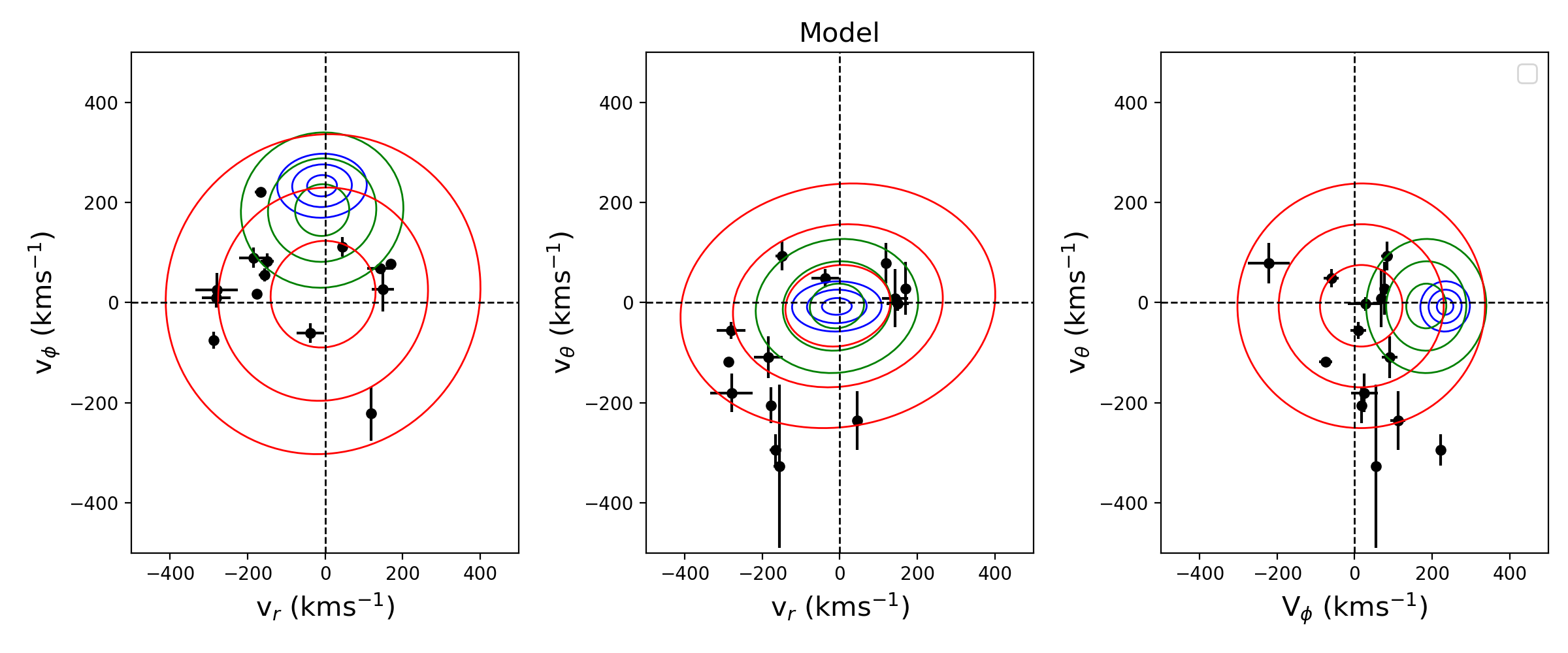}
			\caption{Model velocity ellipsoids using the synthetic white dwarf population within the CFIS footprint from \cite{Fantin2019}. The 1-, 2-, and 3-$\sigma$ velocity ellipsoids for a thin disk (blue), thick disk (green), and halo (red) are shown. The white dwarfs are plotted using distances calculated from the \textit{Gaia} parallaxes (top) and spectroscopic absolute magnitude (bottom) as in Figure \ref{fig:velocities}.}
			\label{fig:model_velocities}
		\end{figure*}

		To emphasize this point, we simulated our selection function using our white dwarf population synthesis code from \cite{Fantin2019}. The velocity distributions were taken from \cite{Robin2017}, who derived them by comparing the Besan\c{c}on model to kinematic data from RAVE and \textit{Gaia} DR1. The resulting 1-, 2-, and 3-$\sigma$ velocity ellipsoids of the thin disk, thick disk, and halo are shown in Figure \ref{fig:model_velocities}. The model values show that all of our white dwarfs lie beyond the 3-$\sigma$ thin disk region, with the majority also falling outside of the 3-$\sigma$ thick disk ellipsoid as well.

		The model can also simulate the number of synthetic white dwarfs that would pass our selection criteria presented in Section \ref{sec:data}. To calculate the probability of a thin disk or thick disk object of lying below the 200 kms$^{-1}$ we ran 10,000 simulations using the parameters presented in \cite{Fantin2019}. The resulting fraction of thin disk objects which would pass our selection method is 3$\times$10$^{-4}$\,\%, while 0.07\,\% of thick disk objects would be selected. Scaling for the mass differences, and given that we observed 13 objects, the expected number of thin and thick disk objects are 0.07 and 1.5 respectively.
		
		It is thus unlikely that all of these objects are simply interloping young thin disk objects and that other factors are responsible for these young objects with halo-like velocities. However, given the large rotational velocity of our highest mass object, J1133+4711, this may be a truly young white dwarf on an extreme disk orbit.
		
		We note that we do expect some thick disk objects to pass our selection criteria, however, given the observed metallicity of thick disk stars, most models of the thick disk assume ages of $\sim$10\,Gyr \citep[see, e.g,][]{Snaith2015, Kilic2017}, which would mean that the currently observable hot white dwarfs descending from this population would still have masses lower than those we calculate.

		\subsection{Class 1: Sample with Accurate Masses}
		\label{sec:class1}
		
		As Section \ref{sec:bias} shows, there likely exists a issue in that the fitting routine that returns higher masses at low SNR unless the mass is quite high. Therefore, we split our sample into three categories: (1) the high SNR objects which pass our original SNR threshold (J0808+4342, J0823+3111, J0842+3338, J1238+3325, J1345+4001, J2328+2138), (2) J1133+4711, which has an SNR below our threshold but has a small uncertainty due to its extreme mass (which is confirmed by the photometric method), and (3) the remaining objects with an SNR at 3900\AA\ below 15. We will perform an analysis on the first category here, before discussing the remaining two in the following section.
		
		\begin{figure*}[!t]
			
			\includegraphics[angle=0,width=\textwidth]{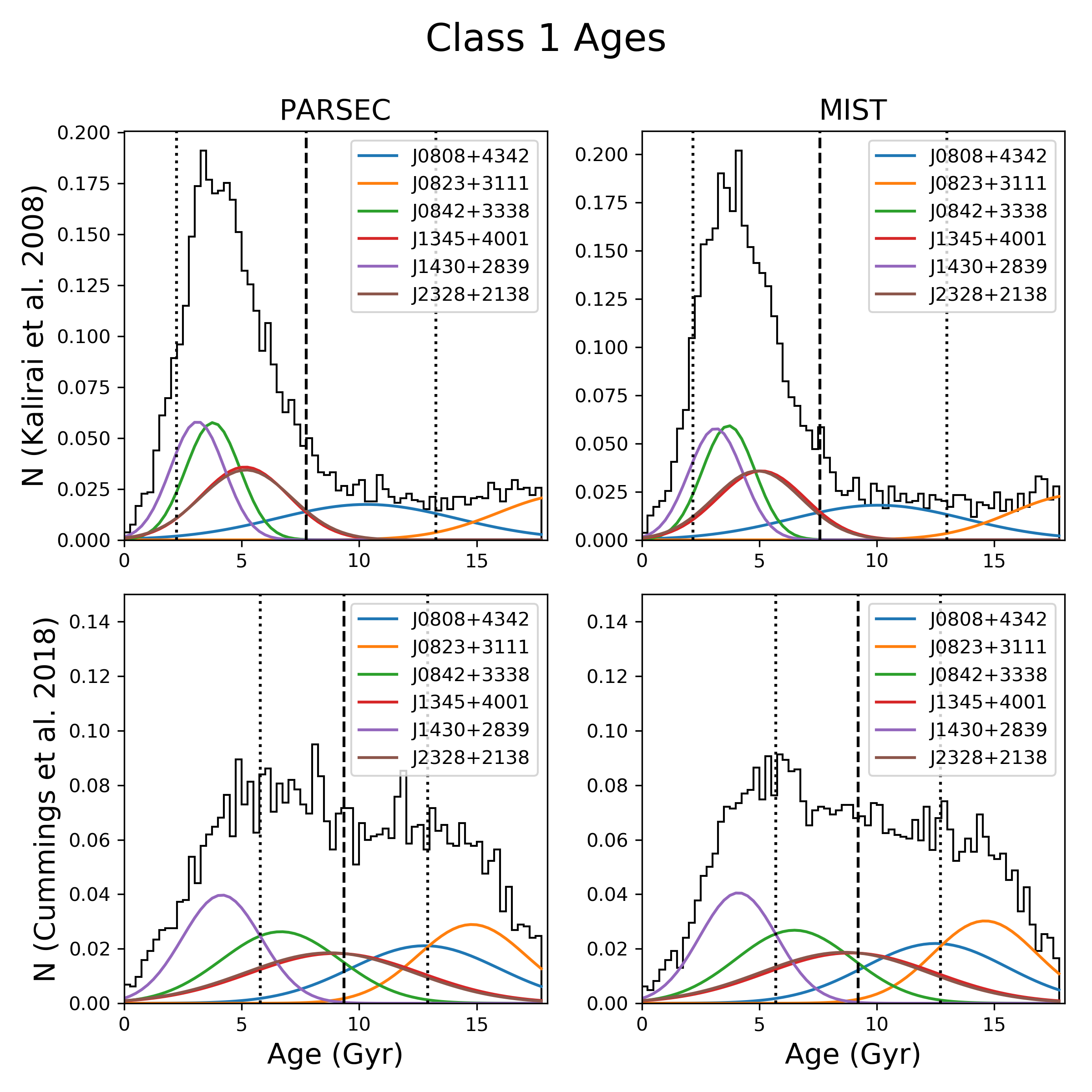}
			\caption{Age histograms for objects belonging to Class 1 (see \ref{sec:class1} for discussion). The \cite{Cummings2018} IFMR returns lower initial masses, resulting in larger overall ages. Also shown are the mean (solid black line), and 1$\sigma$ values (dashed black lines).}
			\label{fig:class1}
		\end{figure*}
		
		In Figure \ref{fig:class1} we present the age histogram for the Class 1 objects. Removing the low SNR objects increases the mean age, while also removing the majority of young objects seen in Figure \ref{fig:ages}. This is particularly true when using the IFMR from \cite{Cummings2018}, which contains a majority of objects with ages at, or above, 10\,Gyr. 
		
		If we only consider the objects which passed our original criteria, the mean mass becomes 0.561\,$\pm$\,0.007, suggesting the halo has a median age of 9.3$\pm$\,1.4\,Gyr using the \cite{Cummings2018} IFMR and MIST isochrones, or 10.8$\pm$\,0.6\,Gyr using the relation from \cite{Kalirai2012}. These results are consistent within the uncertainties with the result of \cite{Kalirai2012}, who derived an age of 11.7\,$\pm$\,0.7\,Gyr, as well as the age of 12.5$^{+1.5}_{-3.4}$\,Gyr from \cite{Kilic2017}.
		
		This sample can also be used to calculate the intrinsic dispersion in age. In order to disentangle the contribution by measurement errors to the measured dispersion we use the method of \cite{Martin2007} and \cite{Collins2013}. This method maximizes the likelihood function,
		
		\begin{equation}
			ML(\mu, \sigma_{\tau}) = \prod_{i=1}^{N} \frac{1}{\sigma_{\mathrm{tot}}}\exp \left [ -\frac{1}{2}\left ( \frac{\mu - \tau_{i}}{\sigma_{\mathrm{tot}}} \right )^2 \right ]
			\label{equation:likelihood}
		\end{equation}
		
		\noindent where $\tau_i$ are the individual ages of each star, $\mu$ is the mean age of the population, and $ \sigma_{\mathrm{tot}} = \sqrt{\sigma_{\tau}^2 + \sigma_i^2}$ is the sum of the dispersion from measurement errors, $\sigma_i$, and the intrinsic dispersion, $\sigma_{\tau}$. 
		
\begin{figure*}[!t]
			
			\includegraphics[angle=0,width=0.49\textwidth]{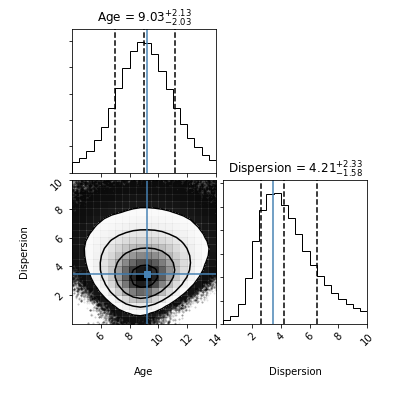}
			\includegraphics[angle=0,width=0.49\textwidth]{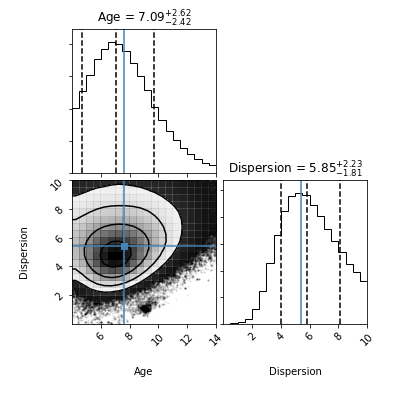}
			\caption{Corner plots for the mean age and intrinsic dispersion in ages for Class 1 objects. (Left): Using the MIST isochrones and \cite{Cummings2018} IFMR. (Right:) Using MIST isochrones and \cite{Kalirai2008} IFMR.}
			\label{fig:corner_halo}
		\end{figure*}

		We use Emcee \citep{Emcee} to maximize the likelihood function and to determine the values and uncertainties associated with the mean and dispersion in equation \ref{equation:likelihood}. The results of simulations with one thousand walkers each taking one million steps can be seen in Figure \ref{fig:corner_halo}. The results remove a burn-in of 100 steps and the three dashed lines show the 16th, 50th, and 84th percentiles of the posterior probability distribution shown in the histograms. The lower and upper uncertainty values presented in the corner plots for the age and dispersion represent the 16th and 84 percentile values respectively.

		Given that changing isochrones did not have a large effect on the total age the simulations use the MIST isochrones and only show a difference based on the choice of IFMR. Using the most up-to-date IFMR from \cite{Cummings2018}, a mean age of 9.03$^{+2.13}_{-2.03}$\,Gyr is measured. The intrinsic dispersion was determined to be 4.21$^{+2.33}_{-1.58}$\,Gyr.
		
		This suggests a more extended star formation history than is typically assumed for this population \citep[$\sim$1\,Gyr. see e.g,][]{Reid2005, Kilic2017}. Recent evidence, however, has shown that the inner halo is more complex and likely dominated by a major accretion event that occurred around 10\,Gyr ago \citep{Helmi2018, Belokurov2018}, allowing for a more extended star formation history in the progenitor. Our ages are also consistent with this event, suggesting that these white dwarfs may have formed as a result of such a merger, or even in the progenitor itself.
		
		\subsection{High Mass Halo White Dwarfs?}

		Given the inconclusive results from Section \ref{sec:bias} regarding our low SNR objects in sample (3), we consider four scenarios for which high-mass white dwarfs on halo-like orbits could occur. The first three scenarios involve Galactic processes: (1) these objects are indeed thin disk objects but have been perturbed somehow onto halo-like orbits (2) these objects may be the end-products of double degenerate mergers; or (3) these objects are indeed old but they were able to prolong their lifetimes through binary interactions. Finally, in a fourth scenario, we consider whether these objects could have been recently accreted from an external galaxy having an extended star-formation history or result from a perturbation from an external galaxy.

		\subsubsection{Scenario I: Ejected Thin Disk Objects}
		
		The first scenario has been proposed to explain the hypothesis of \cite{Oppenheimer2001}, who selected high-velocity white dwarfs and claimed that these objects could be a significant contributor to the unseen matter in the halo. \cite{Davies2002} proposed that thin disk stars could achieve high tangential velocities if they were ejected from a binary system due to a more massive companion experiencing a type II supernova. The star would then evolve, as usual, into a white dwarf while maintaining the high-velocity. The authors simulate a binary interaction for which a 12 M$_{\odot}$ primary star undergoes a supernova resulting in the secondary star being ejected, and produce a relationship between the secondary mass and the kick velocity (see their Figure 3).
		
\begin{figure*}[!t]
			
			\includegraphics[angle=0,width=\textwidth]{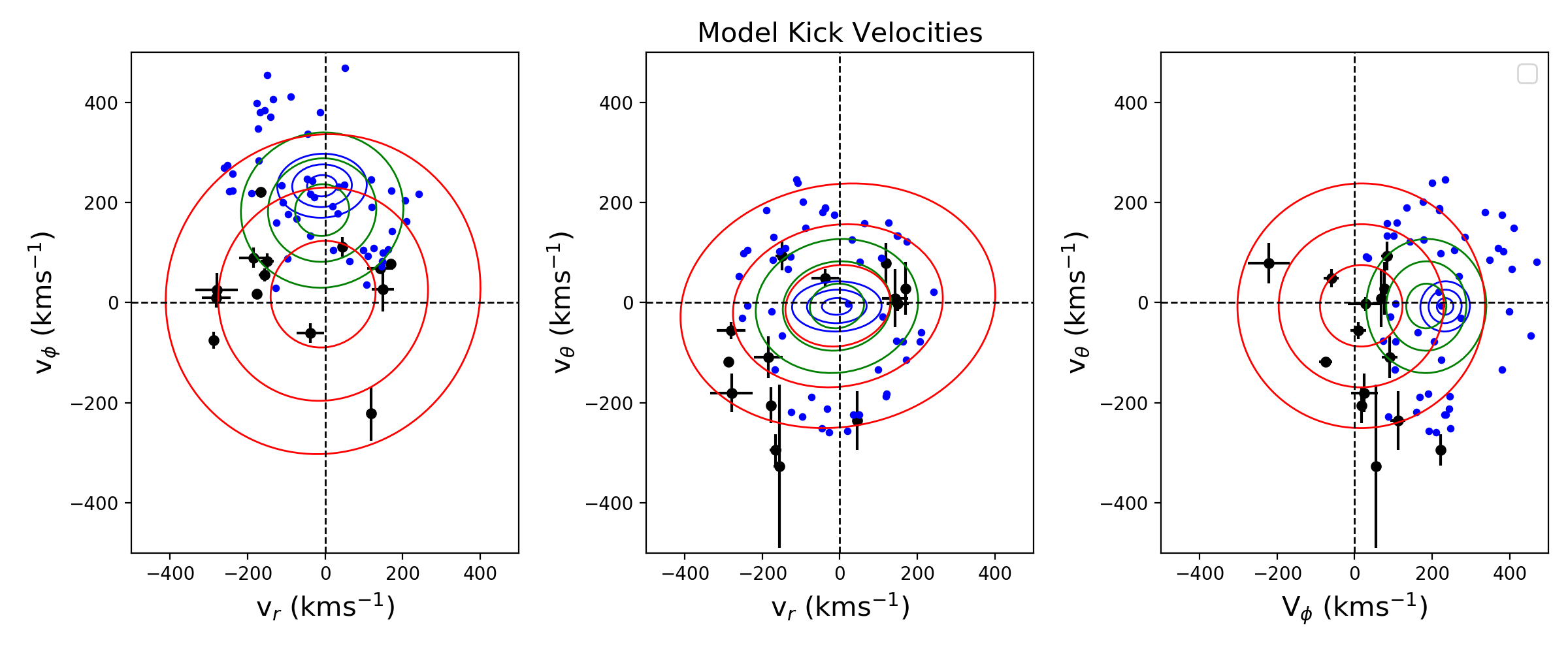}
			\caption{Model velocity ellipsoids using the synthetic white dwarf population within the CFIS footprint from \cite{Fantin2019}. The 1-, 2-, and 3-$\sigma$ velocity ellipsoids for a thin disk (blue), thick disk (green), and halo (red) are shown. The white dwarfs are plotted using distances calculated from spectroscopic absolute magnitude as in Figure \ref{fig:velocities}. White dwarfs which originated in the thin disk and received a kick velocity are plotted in blue.}
			\label{fig:model_velocities_kick}
		\end{figure*}

		To test this hypothesis we select thin disk stars within our white dwarf population synthesis model and impose a kick velocity, selected from their derived relationship to their original velocities. We then apply the same selection function as presented in Section \ref{sec:data} to determine the fraction of ejected objects which would fall within our selected region of the reduced proper motion diagram. The total fraction of thin disk objects which received this kick velocity and passed our selection function is 0.19\,$\pm$0.01\,\%. Thus, for a sample of nearly 30,000 white dwarfs, of which 80\% are likely members of the thin disk, we would detect on the order of 50 objects if all of them received a velocity kick. This can be seen in Figure \ref{fig:model_velocities_kick}. In order to explain our set of high-mass objects we would require that nearly 10\% of all thin disk white dwarfs be ejected during their lifetime, a number that is unrealistic given the lack of high-mass stars produced via the initial-mass function that can produce a type II supernova. 
		
		Furthermore, given that the direction of the kick velocities is likely random, and that the objects are initially rotating in the disk, we expect for this sample of objects to continue to rotate but with a larger velocity dispersion. Our sample of low SNR halo objects does not show any net rotation, and every object lags behind the local standard of rest. This comparison is highlighted in Figure \ref{fig:model_velocities_kick}, where many of the objects which received a kick do not lie in the same region of velocity space as our sample. Thus, we conclude that while objects from the thin disk can be kicked out into the halo, the discrepancy between the velocity distributions seen in the model when compared to our data makes it unlikely that our samples were drawn from the same distribution. To test this, we apply a 2-d Kolmogorov-Smirnov test to our (v$_r$, v$_{\phi}$) data using the method presented in \cite{Fasano1987}. The test returns a p-value of  8.8$\times$10$^{-5}$, suggesting that the velocities are not selected from the same distribution. We thus conclude that this process alone cannot account for the high-mass white dwarfs seen in our sample.

		This scenario may explain J1133+4711, however, as it has a large radial velocity component whilst maintaining a large rotational velocity as well. It also lies close to the 3$\sigma$ contour for the thin disk, suggesting that if it started with a large velocity to begin with, even a minimal kick could allow it to enter a more radial orbit.
		
		\subsubsection{Scenario II: White Dwarf Mergers}

		When calculating the age using isochrones, it is assumed that the object evolved as a single white dwarf without any external influence. Recent observations have shown that the merger of two white dwarfs in a binary system can form a single, more massive white dwarf. This new object would have a mass roughly equivalent to the combined mass of the two stars, although some mass loss may occur. \cite{Temmink2020} found that the resulting merger products have masses higher than white dwarfs formed through single-star evolution, with a mean of 0.71\,M$_{\odot}$.
		
		The merged white dwarf scenario has been used to explain the discrepancy between the binary fraction of main-sequence stars ($\sim$50\%) and the local ($<$25\,pc) white dwarf sample \citep[$\sim$26\%][]{Holberg2016, Toonen2017, Kilic2018}. Furthermore, there exists a high-mass `bump' in the observed mass distribution \citep[see, e.g, Figure 1.4 and][]{Liebert2005, Kleinman2013, Kepler2019}, which can be explained by a double degenerate merger rate of $14 - 25$\%. There are, however, other explanations for this high-mass bump, which include a flattening of the IFMR \citep{ElBadry2018, Cummings2018}. This explanation does not explain the discrepancy in the binary fraction between the local main-sequence stars and the white dwarfs.
		
		\subsubsection{Scenario III: Descendants of Blue Stragglers}
		
		Figure \ref{fig:initial_mass} shows that many of the objects have progenitor masses between 1.5 and 2 M$_{\odot}$, which are typical of A-type stars. Despite having main-sequence lifetimes on the order of 1 Gyr \citep{Hurley2000, PARSEC, Choi2016}, these stars are still seen in old populations like globular clusters \citep[see, e.g,][]{Harris1993, Parada2016}, as well as in the thick disk and halo of the Milky Way \citep{Santucci2015} and are called blue stragglers (BSs). These objects are thought to form as a result of mass-transfer binaries or direct mergers of two stars \citep{Stryker1993}. The lifetimes have been estimated to be $200 - 300$\,Myr once formed, and their post-main-sequence evolution is the same as a star with the same initial masses \citep{Parada2016}. Thus, halo BSs would create higher mass white dwarfs with kinematics consistent with those seen in our sample.
		
		To study this scenario, we require a sample of metal-poor, main-sequence, A-type stars in the solar neighbourhood on halo-like orbits. \cite{Preston1994} used \textit{UBV} photometry to select blue stars with [Fe/H]$<-$1.0, (dubbed blue metal-poor; BMP) stars, within 2\,kpc of the Sun and found a space density of 450$^{+300}_{-150}$,kpc$^{-3}$. Follow-up spectroscopy presented in \cite{Wilhelm1999}, however, showed that a large fraction of these objects ($\sim$50\%) were more metal-rich ([Fe/H]$>-$1.0) than previously believed, suggesting that many of these objects may not be from an old population.

		A sub-sample of the BMP stars from \cite{Preston1994} were monitored by \cite{Preston2000} to test their binarity through radial velocity variations. Their results suggest that more than half of these objects were in a binary system, suggesting that they were indeed blue stragglers formed through mass accretion. Thus, if half of the original sample were true blue stragglers, and half of these were indeed metal-poor as suggested by \cite{Wilhelm1999}, the space density of these blue stragglers would be 1.1$^{+0.75}_{-0.38}\times$10$^{-7}$\,pc$^{-3}$.

		The space density for young, hot, halo objects (T$_{\textrm{eff}} >$ 10,000\,K) studied in this work has been calculated to be 1.2$\times$10$^{-6}$\,pc$^{-3}$ using the luminosity function of \cite{Munn2017}, a factor of 10 larger than the space density of halo blue stragglers. This is likely an upper limit, as many of these BMPs may not be part of the in-situ halo given their rotational velocity of 128 kms$^{-1}$ presented by \cite{Preston1994}, which lies intermediate to the thick disk and halo. Thus, we conclude that this scenario alone can not account for the observed high mass halo white dwarfs, although it is worth considering given the recent work by \cite{Parada2016} which showed that blue stragglers evolve off the main-sequence as if they had always been at their current mass, suggesting they will form higher mass white dwarfs. Our upper limit of 9$^{+6}_{-3}$\% contamination is also consistent with the increased fraction of AGB stars they observe in 47 Tucanae that they attribute to evolved blue stragglers.
		
		\subsubsection{Scenario 4: Accretion or Perturbation from a Satellite}
		
		\cite{Preston1994} suggested that many of these BMP stars could be of extragalactic origin as their young ages and low metallicity could occur in smaller dwarf galaxies seen surrounding the Milky Way. If one of these objects was accreted recently it could explain the origin of such objects and their resulting white dwarfs. The issue with this scenario is that, given the inferred progenitor masses, these objects were formed within the past 2\,Gyr regardless of the progenitor metallicity. Combining this with the large fraction of our sample means that we would require a large merger event within the past few Gyr, which to our knowledge has not been reported within the literature \citep{Naab2017, Ruiz2020}.

		Much of the recent focus has been on the Gaia-Enceladus merger event \citep{Helmi2018, Belokurov2018}, which is thought to be the last major merger event in the Milky Way. This event, however, is thought to have occurred more than 10\,Gyr ago, and could be used to explain the formation of the thick disk \citep{Helmi2018}. Thus, a major merger could explain the formation of our low-mass objects, however, there is a lack of evidence for a major merging event over the last few Gyr \citep{Naab2017, Ruiz2020}.

		There is, however, evidence for interactions with more minor satellites over the past few Gyr, including the Sagittarius dwarf galaxy. \cite{Ruiz2020} studied the local 2\,kpc bubble and found signatures of an increased star-formation rate at three separate epochs: 5.7\,Gyr, 1.9\,Gyr and 1.0\,Gyr, the last two of which are consistent with the ages derived in Section \ref{sec:Results}. These bursts of star formation coincide with the proposed pericentre passages of the Sagittarius dwarf galaxy and are visible in both the thin and thick components of the Milky Way. \cite{Ruiz2020} conclude that these episodes could be explained by the interaction between the Milky Way disk and Sagittarius, however, the exact physical mechanism has yet to be explained. Thus, the high-mass stars formed in the thick disk during one of these episodes could form the high-mass white dwarfs observed in our sample, although more studies are needed to determine the orbits of the objects formed during these events.

		\section{Summary and Conclusions}
		\label{sec:conclusions}

		This paper has presented follow-up spectroscopy for 18 candidate halo white dwarfs performed at the twin 8-m Gemini Observatories. Our results are as follows:
		
		\begin{itemize}
			\item We observe 13 DA, 2 DC, 1 DZ, and 2 peculiar white dwarfs in our sample, representing a large variation in types relative to previous samples.
			
			\item The discovery of two peculiar, but rare, white dwarfs reinforces the conclusion of \cite{Raddi2019} who suggested that these objects could form as a result of a peculiar class of supernovae occurring in binary systems that ejects the white dwarf, thus increasing its velocity.
			
			\item For the sample of DA white dwarfs, we calculate their temperature and mass by fitting the Balmer series. The results show a peak at 0.58\,M$_{\odot}$ with a number of outliers. 
			
			\item We convert the white dwarf mass to a total age using the initial-to-final mass relation of \cite{Kalirai2008} and \cite{Cummings2018} in combination with MIST and PARSEC stellar isochrones. The ages show a large spread, which is not expected for a halo sample.
			
			\item We investigate whether the model fitting routine favors higher white dwarf masses at lower signal-to-noise and find that the model predicts higher masses for objects with lower signal-to-noise ratios at H$_8$. We attempt to study this situation in more detail by fitting a combination of sub-exposures for the low-mass, high SNR objects, however, the results are inconclusive. We recommend SNR ratios above 15 at H$_{8}$ for future large scale studies.
			
			\item Given the uncertainty on the masses at low SNR, we split the white dwarfs into three classes: (1) Those with SNR $>$ 15 at H$_8$ as recommended by \cite{Kepler2006}, (2) J1133+4711 which has a mass of 1.22 M$_{\odot}$ in both the spectroscopic and photometric methods, and (3) the remaining objects with SNR $<$ 15. 
			
			\item We find that the high SNR objects have a mean age of 9.3$\pm$\,1.4\,Gyr using the \cite{Cummings2018} IFMR and MIST isochrones, or 10.8$\pm$\,0.6\,Gyr using the relation from \cite{Kalirai2012}. These results are consistent with previous studies of the local halo population.
			
			\item We determine the intrinsic dispersion in the local halo ages for Class (1) to be 4.21$^{+2.33}_{-1.58}$\,Gyr using a maximum likelihood method. This suggests a more extended star formation history for the local halo than is typically assumed.

			\item We investigate four scenarios that would produce high-mass white dwarfs on halo-like orbits, such as those in class (2) and (3). Using our Milky Way white dwarf population synthesis code we find that ejected thin disk objects alone can not explain the sample, however, a combination of white dwarf mergers, halo blue stragglers, or increased star formation as a result of an interaction with the Sagittarius dwarf galaxy could explain this sample.
			
		\end{itemize}

		This work highlights the need for high signal-to-noise data, particularly at H$_8$, when measuring white dwarf masses as low-SNR data may induce a bias in the resulting white dwarf masses. This is particularly true around 12,000\,K where the Balmer lines are at their maximum equivalent widths. We show that the average age of low SNR objects is between 1 and 3 Gyr, whereas when we exclude the objects with SNR below the recommendation from \cite{Kepler2006}, we find the age of the inner halo to be 9.3$\pm$\,1.4\,Gyr using the \cite{Cummings2018} IFMR and MIST isochrones, or 10.8$\pm$\,0.6\,Gyr using the relation from \cite{Kalirai2012}, which are consistent with previous results for the inner halo.
		
		We note, however, that we can not rule out these objects as truly being high-mass white dwarfs belonging to the Galactic halo. Blue stragglers, double degenerate mergers, binary interactions, or enhanced star formation in the thick disk due to an interaction with a dwarf galaxy can all produce high-mass white dwarfs with halo-like velocities. Future large spectroscopic surveys such as WEAVE \citep{WEAVE}, 4MOST \citep{4MOST}, and the Mauna Kea Spectroscopic Explorer \citep{MSE} will be needed to increase the spectroscopic sample of halo white dwarfs to accurately determine their mass and age distribution.
		
		This work has also reiterated the importance of the IFMR. In particular, much work still needs to be done at the low-mass end where halo stars are currently producing white dwarfs \citep[for a detailed discussion see][]{Fantin2020}. Future large observatories such as the Mauna Kea Spectroscopic Observatory and the extremely large telescopes (the Thirty Meter Telescope, the Giant Magellan Telescope, and the Extremely Large Telescope) will be needed to acquire a large sample of spectroscopic masses in globular clusters given their large distances and faint populations. In addition to the ground-based observatories, recent work by \cite{Gentile2019} suggests that the James Webb Space Telescope can also be used by targeting the Paschen series in the near-infrared. Combining these observatories with the previously mentioned spectroscopic surveys will provide an unprecedented look into the formation and evolution of the Galactic inner halo, and may push studies beyond the Solar neighborhood.

		\acknowledgements
		
		Based on observations obtained at the Gemini Observatory (GN-2018B-Q-115, GN-2019A-Q-224 \& GS-2019A-Q-227), and processed using the Gemini PyRAF package, which is operated by the Association of Universities for Research in Astronomy, Inc., under a cooperative agreement with the NSF on behalf of the Gemini partnership: the National Science Foundation (United States), National Research Council (Canada), CONICYT (Chile), Ministerio de Ciencia, Tecnología e Innovaci{\'o}n Productiva (Argentina), Ministério  Ciência, Tecnologia e Inovação (Brazil), and Korea Astronomy and Space Science Institute (Republic of Korea).
		
		GT acknowledges support from the Agencia Estatal de Investigaci{\'o}n (AEI) of the Ministerio de Ciencia e Innovaci{\'o}n (MCINN) under grant with reference (FJC2018-
		037323-I).
		
		This work is based on data obtained as part of the Canada France Imaging Survey, a CFHT large program of the National Research Council of Canada and the French Centre National de la Recherche Scientifique. Based on observations obtained with MegaPrime/MegaCam, a joint project of CFHT and CEA Saclay, at the Canada France Hawaii Telescope (CFHT) that is operated by the National Research Council (NRC) of Canada, the Institut National des Science de l'Univers (INSU) of the Centre National de la Recherche Scientifique (CNRS) of France, and the University of Hawaii. This research used the facilities of the Canadian Astronomy Data Centre operated by the National Research Council of Canada with the support of the Canadian Space Agency. 
		
		The Pan-STARRS1 Surveys (PS1) and the PS1 public science archive have been made possible through contributions by the Institute for Astronomy, the University of Hawaii, the Pan-STARRS Project Office, the Max-Planck Society and its participating institutes, the Max Planck Institute for Astronomy, Heidelberg and the Max Planck Institute for Extraterrestrial Physics, Garching, The Johns Hopkins University, Durham University, the University of Edinburgh, the Queen's University Belfast, the Harvard-Smithsonian Center for Astrophysics, the Las Cumbres Observatory Global Telescope Network Incorporated, the National Central University of Taiwan, the Space Telescope Science Institute, the National Aeronautics and Space Administration under Grant No. NNX08AR22G issued through the Planetary Science Division of the NASA Science Mission Directorate, the National Science Foundation Grant No. AST-1238877, the University of Maryland, Eotvos Lorand University (ELTE), the Los Alamos National Laboratory, and the Gordon and Betty Moore Foundation.
		
		This work has made use of data from the European Space Agency (ESA) mission
		{\it Gaia} (\url{https://www.cosmos.esa.int/gaia}), processed by the {\it Gaia}
		Data Processing and Analysis Consortium (DPAC,
		\url{https://www.cosmos.esa.int/web/gaia/dpac/consortium}). Funding for the DPAC
		has been provided by national institutions, in particular the institutions
		participating in the {\it Gaia} Multilateral Agreement.

	\end{document}